\documentclass[useAMS,usenatbib]{mn2e}
\usepackage{graphicx}

\def\be{\begin{equation}} 
\def\ee{\end{equation}}

\def\HI{\hbox{H~$\scriptstyle\rm I\ $}}

\def\gsim{\lower.5ex\hbox{\gtsima}} 
\def\lsim{\lower.5ex\hbox{\ltsima}} \def\gtsima{$\; \buildrel > \over 
\sim \;$} \def\ltsima{$\; \buildrel < \over \sim \;$} \def\prosima{$\; 
\buildrel \propto \over \sim \;$} \def\gsim{\lower.5ex\hbox{\gtsima}} 
\def\lsim{\lower.5ex\hbox{\ltsima}} 
\def\simgt{\lower.5ex\hbox{\gtsima}} 
\def\simlt{\lower.5ex\hbox{\ltsima}} 
\def\simpr{\lower.5ex\hbox{\prosima}} \def\la{\lsim} \def\ga{\gsim} 
  
 \def\gtsima{$\; \buildrel > \over \sim \;$} 
\def\ltsima{$\; \buildrel < \over \sim \;$} 
\def\gsim{\lower.5ex\hbox{\gtsima}} 
\def\lsim{\lower.5ex\hbox{\ltsima}} 
\def\simgt{\lower.5ex\hbox{\gtsima}} 
\def\simlt{\lower.5ex\hbox{\ltsima}} 
\def\simpr{\lower.5ex\hbox{\prosima}} 
\def\la{\lsim} 
\def\ga{\gsim}

\def\E3{{\cal E}_{\rm g}^{III}}

\def\Msun{\rm M_\odot}
\def\Zsun{\rm Z_\odot}
\def\cmpc{\rm cMpc}
\def\kpc{\rm Kpc}
\def\Msun{\rm M_\odot}
\def\Zsun{\rm Z_\odot}
\def\M*{M_*}
\def\Z*{Z_*}
\def\L*{L_*}

\title[Assembly of high-z galaxies]{Simulating the assembly of galaxies at redshifts ${\bf z = 6 - 12}$} 
\author[Dayal et al.]{Pratika Dayal$^{1}$\thanks{E-mail:prd@roe.ac.uk (PD)}, James S Dunlop$^{1}$, Umberto Maio$^{2,3}$ \& Benedetta Ciardi$^{4}$  \\ 
$^{{1}}$ SUPA\thanks{Scottish Universities Physics Alliance}, Institute for Astronomy, University of Edinburgh, Royal Observatory, Edinburgh, EH9 3HJ, UK \\
$^{2}$ Max-Planck-Institut f\"ur extraterrestrische Physik Giessenbachstrasse 1, 85748 Garching, Germany \\
$^{3}$ INAF - Osservatorio Astronomico di Trieste, via  G. B. Tiepolo 11, 34131 Trieste, Italy \\
$^{4}$ Max Planck Institut f\"ur Astrophysik, Karl-Schwarzschild-Strasse 1, 85741 Garching, Germany }

\begin{document} 
 
\date{} 
 
 
\maketitle 
 
\label{firstpage} 
\begin{abstract} 

We use state-of-the-art simulations to explore the physical evolution of galaxies in the first billion years of cosmic time. First, we demonstrate that our
model reproduces the basic statistical properties of the observed Lyman-break galaxy (LBG) population at $z = 6 - 8$, including
the evolving ultra-violet (UV) luminosity function (LF), the stellar-mass density ($SMD$), and the average specific star-formation rates ($sSFR$) of LBGs with $M_{UV} < -18$ (AB mag). Encouraged by this success we present predictions for the behaviour of fainter LBGs extending down to
$M_{UV} \simeq -15$ (as will be probed with the {\it James Webb Space Telescope}) and have interrogated our simulations to try to gain insight
into the physical drivers of the observed population evolution. We find that mass growth due to star formation in the mass-dominant progenitor
builds up about 90\% of the total $z\sim 6$ LBG stellar mass, dominating over the mass contributed by merging throughout this era. Our simulation suggests that the apparent ``luminosity evolution" depends on the luminosity range probed: the steady brightening of the bright end of the LF is driven primarily by genuine physical luminosity evolution and arises due to a fairly steady increase in the UV luminosity (and hence star-formation rates) in the most massive LBGs; e.g. the progenitors of the $z \simeq 6$ galaxies with $M_{UV} < -18.5$ comprised $\simeq 90$\% of the galaxies with $M_{UV} < -18$ at $z \simeq 7$, and $\simeq 75$\% at $z \simeq 8$. However, at fainter luminosities the situation is more complex, due in part to the more stochastic star-formation histories of lower-mass objects; the progenitors of a significant fraction of $z \simeq 6$ LBGs with $M_{UV} > -18$ were in fact brighter at $z \simeq 7$ (and even at $z \simeq 8$) despite obviously being less massive at earlier times. At this end, the evolution of the UV LF involves a mix of positive and negative luminosity evolution (as low-mass galaxies temporarily brighten then fade) coupled with both positive and negative density evolution (as new low-mass galaxies form, and other low-mass galaxies are consumed by merging). We also predict the average $sSFR$ of LBGs should rise from $sSFR \simeq 4.5\,{\rm Gyr}^{-1}$ at $z \simeq 6$ to $sSFR \simeq 11\,{\rm Gyr}^{-1}$ by $z \simeq 9$.

\end{abstract}


\section{Introduction}
\label{intro}
In the standard Lambda Cold Dark Matter ($\Lambda$CDM) scenario, the first structures to form in the Universe were low-mass 
dark-matter (DM) halos, and these building blocks merged hierarchically to form successively larger structures 
\citep[e.g.][]{blumenthal1984}. While the initial conditions (at $z \approx 1100$) for the formation of such structures 
are well established as being nearly scale-invariant, small-amplitude density fluctuations \citep[e.g.][]{komatsu2009}, 
until recently there has been relatively little direct observational information on the emergence of the first 
galaxies. These early galaxies changed the state of the intergalactic medium 
(IGM) from which they formed: they polluted it with heavy elements and heated and (re)ionized it, thereby affecting 
the evolution of all subsequent generations of galaxies \citep{barkana-loeb2001, ciardi-ferrara2005, robertson2010}. 

As recently reviewed by \citet{dunlop2012b}, until the advent of sensitive near-infrared imaging with the installation 
of Wide Field Camera 3 (WFC3/IR) on the {\it Hubble Space Telescope} (HST) in 2009, there existed at most one convincing 
galaxy candidate at redshifts $z > 7$ \citep{bouwens2004}. Now, however, over one hundred galaxies have been uncovered with HST in the redshift 
range $6.5 < z < 8.5$ \citep[e.g.][]{oesch2010, bouwens2010a, mclure2010, finkelstein2010, mclure2011, lorenzoni2011, bouwens2011, oesch2012}, 
and the new generation of ground-based wide-field near-infrared surveys, such as UltraVISTA \citep{mccracken2012}
are reaching the depths required to reveal more luminous galaxies at $z \simeq 7$ \citep{castellano2010a, ouchi2010, bowler2012}.

The assembly of significant samples of galaxies within $< 1$\,Gyr of the big bang has enabled the first meaningful determinations
of the form and evolution of the UV LF of galaxies over the redshift range $6 < z < 8$,
reaching down to absolute magnitudes $M_{UV} \simeq -18$ \citep{mclure2010, oesch2010, finkelstein2010,
bouwens2011, finkelstein2012b, bradley2012}. It has also facilitated the study of the typical properties of these young galaxies, 
both through follow-up spectroscopy \citep{schenker2012, curtis-lake2012a}, and from analysis of their average 
broad-band (HST + {\it Spitzer}) spectral energy distributions (SEDs) \citep[e.g.][]{labbe2010a, bouwens2010, gonzalez2010, 
finkelstein2010, wilkins2011, mclure2011, dunlop2012, bouwens2012, finkelstein2012a, curtis-lake2012b, stark2012, labbe2012}.  

The results of these studies are still a matter of vigorous debate, and will undoubtedly be refined substantially 
over the next few years, as the HST/{\it Spitzer} and ground-based near-infrared imaging database expands and improves, 
and deep multi-object near-infrared spectroscopy becomes routine. Nevertheless, the current situation can be broadly
summarized as follows. First, there is good agreement over the basic form and evolution of the galaxy UV LF at $z \simeq 6, 7, 8$, albeit the current data do not allow the degeneracy between the Schechter function parameters
to be unambiguously broken. Second, there is as yet no meaningful information on the number density of galaxies
at $z > 8.5$. Third, there is some tentative evidence for a drop in the prevalence of Lyman-$\alpha$ line emission at $z \simeq 7$,
as arguably expected from an increasingly neutral ISM. Fourth, various studies suggest that the specific star formation rate 
($sSFR =$ star-formation rate/stellar mass) of star-forming galaxies, after rising from $z = 0$ to $z \simeq 3$, then remains remarkably constant
out to $z \simeq 7$. Fifth, there is growing evidence that this last result may need to be modified in the light of 
increasingly strong contributions from nebular line emission with increasing redshift. Sixth, the UV continua of the highest redshift 
galaxies is fairly blue (indicative of less dust than at lower redshifts), but there is as yet no evidence for exotic stellar populations, as might be revealed by extreme UV slopes. 

Despite the remaining uncertainties, it is clear that these new observational results already provide an excellent test-bed 
for the increasingly complex galaxy formation and evolution models/simulations that have recently 
been developed in a number of studies \citep[e.g.][]{finlator2007, dayal2009, dayal2010a, salvaterra2011, dayal2011a, forero2011, dayal2012, yajima2012}. 

In this study, our aim is not only to test the ability of the latest galaxy formation models to reproduce 
these new observational results at high-redshift, but also to use our simulations to obtain a better physical understanding 
of how the emerging galaxies evolve in luminosity and grow in stellar mass. 
We choose $z \simeq 6$ as the baseline, and explore the evolution and 
physical properties of the progenitors of $z\simeq 6$ galaxies out to $z \simeq 12$.
We explore the properties of the currently observable galaxies with $M_{UV} > -18$, but also present predictions reaching
over an order-of-magnitude fainter, to $M_{UV} \simeq -15$. Such faint galaxies are of interest because 
i) they are extremely numerous, and thus provide good number statistics for 
observational predictions, ii) they are expected to have low dust content \citep[see]{dayal2012}, 
simplifying the predictions for rest-frame UV observations, 
iii) current evidence suggest it is these faint galaxies which likely provide the bulk of the photons responsible 
for Hydrogen reionization \citep{choudhury-ferrara2007,salvaterra2011}, iv) such galaxies are within reach of 
planned observations with the {\it James Webb Space Telescope} (JWST) at the end of the decade, 
and v) the simulation of such faint dwarf galaxies requires a high mass-resolution, achievable with the simulation utilised in this work.
Specifically, we use state-of-the-art simulations with a box size of $10\, \cmpc$ (comoving Mpc), and include molecular hydrogen cooling, a 
careful treatment of metal enrichment and of the transition from Pop III to Pop II stars, 
along with modelling of supernova (SN) feedback, as explained in Sec. \ref{simulations} that follows. 

\section{Cosmological simulations}
\label{simulations}
The simulation used in this work is now briefly described, and interested readers are referred to 
\citet{maio2007,maio2009,maio2010} and \citet{campisi2011} for complete details. 

\subsection{Simulation description}
\label{sim_des}

We use a smoothed particle hydrodynamic (SPH) 
simulation carried out using the TreePM-SPH code {\small {GADGET-2}} \citep{springel2005}. The cosmological model corresponds 
to the $\Lambda$CDM Universe with DM, dark energy and baryonic density parameter values of 
($\Omega_{\rm m },\Omega_{\Lambda}, \Omega_{\rm b}) = (0.3,0.7,0.04)$, a Hubble constant $H_0 = 100h= 70 {\rm km\, s^{-1} Mpc^{-1}}$, 
a primordial spectral index $n_s=1$ and a spectral normalisation $\sigma_8=0.9$. The periodic simulation box has a 
comoving size of $10h^{-1} \cmpc$. It contains $320^3$ DM particles and, initially, an equal number of gas particles. 
The masses of the gas and DM particles are $3\times 10^5 h^{-1} \Msun$ and $2 \times 10^6 h^{-1} \Msun$, 
respectively. The maximum softening length for the gravitational force is set to $0.5 h^{-1}\kpc$ and the value of the smoothing length for the SPH kernel for the computation of hydrodynamic forces is allowed to drop at most to half of the gravitational softening.

The code includes the cosmological evolution of ${\rm e^-}$, ${\rm H}$, ${\rm H^-}$, ${\rm He}$, ${\rm He^+}$, 
${\rm He^{++}}$, ${\rm H_2}$, ${\rm H^+_2}$, ${\rm D}$, ${\rm D^+}$, ${\rm HD}$, ${\rm HeH^+}$ 
\citep{yoshida2003, maio2007, maio2009}, and gas cooling from resonant and fine-structure lines \citep{maio2007} . The runs track individual heavy elements (e.g. C, O, Si, Fe, Mg, S), and the transition from the metal-free PopIII to the metal-enriched PopII/I regime is determined by the underlying 
metallicity of the medium, ${\rm Z}$, compared with the critical value of ${\rm Z_{crit}} = 10^{-4} \Zsun$ 
\citep[and references therein]{bromm-loeb2003}, but see also \citet{schneider2003,
schneider2006} who find a much lower value of ${\rm Z_{crit}} = 10^{-6} \Zsun$. However, the precise 
value of $\rm Z_{crit}$ is not crucial; the large metal yields of the first stars easily pollute the medium to 
metallicity values above $\simeq 10^{-4}-10^{-3}~\Zsun$ \citep{maio2010, maio2011}. If ${\rm Z<Z_{crit}}$, a \citet{salpeter1955} 
 initial mass function (IMF) is used, with a mass range of $100-500\, \Msun$ \citep{tornatore2007b}; otherwise, a standard 
Salpeter IMF is used in the mass range $0.1-100\,\Msun$, and a SNII range is used for $8-40\,\Msun$ \citep{bromm2009, maio2011}. The runs include the star formation prescriptions of \citet{springel-hernquist2003b} such that the interstellar medium (ISM) is described as an ambient hot gas containing cold clouds, which provide the reservoir for star formation, with the two phases being in pressure equilibrium; each gas particle can spawn at most 4 star particles. The density of the cold and of the hot phase represents an average over small regions of the ISM, within which individual molecular clouds cannot be resolved by simulations sampling cosmological volumes. The runs also include the feedback model described in \citet{springel-hernquist2003b} which includes thermal feedback (SN injecting entropy into the ISM, heating up nearby particles and destroying molecules), chemical feedback \citep[metals produced by star formation and SN are carried by winds and pollute the ISM over $\sim$\,kpc scales at each epoch as shown in][]{maio2011} and mechanical feedback (galactic winds powered by SN). In the case of mechanical feedback, the momentum and energy carried away by the winds are calculated assuming that the galactic winds have a fixed velocity of $500 \, {\rm km \, s^{-1}}$ with a mass upload rate equal to twice the local star formation rate \citep{martin1999}, and carry away a fixed fraction ($25\%$) of the SN energy (for which the canonical value of $10^{51} \,{\rm ergs}$ is adopted).

Further, the run assumes a metallicity-dependent radiative cooling \citep{sutherland-dopita1993, maio2007} and a uniform redshift-dependent UV Background (UVB) produced by quasars and galaxies as given by \citet{haardt-madau1996}. The chemical model follows the detailed stellar evolution of each SPH particle. 
At every timestep, the abundances of different species are consistently derived 
using the lifetime function \citep{padovani-matteucci1993} and metallicity-dependant stellar yields; 
the yields from SNII, AGB stars, SNIa and pair instability SN (PISN; for primordial SN nucleosynthesis) 
have been taken from \citet{woosley-weaver1995}, \citet{vandenhoek1997}, 
\citet{thielemann2003} and \citet{heger-woosley2002}, respectively. Metal mixing is mimicked by smoothing 
metallicities over the SPH kernel. 

Galaxies are recognized as gravitationally-bound groups of at least 32 total (DM+gas+star) particles 
by running a friends-of-friends (FOF) algorithm, with a comoving linking-length of 0.2 in units of the 
mean particle separation. Substructures are identified by using the {\small SubFind} algorithm 
\citep{springel2001,dolag2009} which discriminates between bound and non-bound particles. When compared to the standard Sheth-Tormen mass function \citep{sheth-tormen1999}, the simulated mass function is complete for halo masses $M_h \geq 10^{8.5} \, \Msun$ in the entire redshift range of interest, i.e. $z \simeq 6-12$; galaxies above this mass cut-off are referred to as the ``complete sample''. Of this complete sample, at each redshift of interest, we identify all the ``resolved" galaxies that contain a minimum of $4N$ gas particles, where $N=32$ is the number of nearest neighbours used in the SPH computations. We note that this is twice the standard value of $2N$ gas particles needed to obtain reasonable and converging results \citep{navarro-white1993, bate-burkert1997}. Since this work concerns the buildup of stellar mass, at any redshift, we impose an additional constraint and only use those resolved galaxies that contain at least 10 star particles so as to get a statistical estimate of the composite SED. For each resolved galaxy used in our calculations (with $M_h \geq 10^{8.5} \, \Msun$, more than $4N$ gas particles and a minimum of 10 star particles) we obtain the properties of all its star particles, including the redshift of, and mass/metallicity at formation; we also compute global properties including the total stellar/gas/DM mass, mass-weighted stellar/gas metallicity and the instantaneous star-formation rate (SFR). 

We caution the reader that while all the calculations in this paper have been carried out galaxies that contain a minimum of 10 star particles, increasing this limit is likely to affect the selection of the least massive objects. To quantify this effect, we have carried out a brief resolution study, the results of which are shown in Fig. \ref{res_uvlf}. While increasing the selection criterion to a minimum of 20 star particles does not affect the UV LFs at any redshift, increasing this value to 50 star particles only leads to a drop in the number of galaxies in the narrow range between $M_{UV}=-15 $ and $-15.5$ for $z \simeq 7-12$; with the drop increasing with increasing redshift.   

We now briefly digress to discuss some of the caveats involved in the simulation and interested readers are referred to \citep{tornatore2007a, maio2009, maio2010, campisi2011} for complete details: first, the simulation has been run using a star formation density threshold of $7\, {\rm cm^{-3}}$. Obviously, the higher this threshold, the later is the onset of star formation as the gas needs more time to condense. However, exploring density thresholds ranging between $0.1-70 \, {\rm cm^{-3}}$, \citet{maio2010} have shown that the the SFRs obtained from all these runs converge as early as $z \sim 13$. Second, although the simulation uses a Salpeter IMF, as of now, there is no strong consensus on whether the IMF is universal and time-independent or whether local variations in temperature, pressure and metallicity can affect the mass distribution of stars; indeed, \citet{larson1998} claim that the IMF might have been more top-heavy at higher redshifts. The exact form of the IMF for metal-free stars is even more poorly understood, and remains a matter of intense speculation \citep[e.g.][]{larson1998, nakamura-umemura2001, omukai-palla2003}. However, changing the slope of the IMF for PopIII stars between $-1$ and $-3$ does not affect results regarding the metallicity evolution or the SFR as the fraction of PISN is always about $0.4$ \citep{maio2010}. Third, we have used a mass range of $100-500\, \Msun$ for PopIII stars. While many studies suggest the IMF of PopIII stars to be top heavy \citep[e.g.][]{larson1998, abel2002, yoshida2004}, there is also evidence for a more standard PopIII IMF with masses well below $\simeq 100 \, \Msun$ \citep[e.g.][]{yoshida2007, suda-fujimoto2010, greif2011}. The main difference between these two scenarios is that the $100-500\, \Msun$ IMF yields very short PopIII lifetimes which pollute the surrounding medium in a much shorter time compared to the case of the normal IMF; as expected, the top heavy IMF naturally results in an earlier PopIII-II transition which directly affects the metal pollution and star formation history \citep{maio2010}. Fourth, \citet{tornatore2007a} have studied the effects of different stellar lifetime functions and found that the \citet{maeder-meynet1989} function differs from the \citet{padovani-matteucci1993} function used here for stars less than $1 \, \Msun$; although the SNII rate remains unchanged, the SNIa rate is shifted to lower redshifts for the former lifetime function. Fifth, although we have used a modified  \citep{sutherland-dopita1993} cooling function so that it depends on the the abundance of different metal species, due to a lack of explicit treatment of the metal diffusion, spurious noise may arise in the cooling function due to a heavily enriched gas particle being next to a metal-poor particle. To solve this problem, while each particle retains its metal content, the cooling rate is computed by smoothing the gas metallicity over the SPH kernel. Sixth, ejection of both gas and metal particles into the IGM remains a poorly understood astrophysical process. Although in our simulations, winds originate as a result of the kinetic feedback from stars, scenarios like gas stripping, shocks and thermal heating from stellar processes could play an important role, given that it is extremely easy to lose baryonic content from the small DM potential wells of high-$z$ galaxies. 

Finally, although a number of earlier works have used either pure DM \citep[e.g.][]{mcquinn2007, iliev2012} or SPH \citep[e.g.][]{finlator2007, dayal2009, dayal2010a, nagamine2010, zheng2010, forero2011} cosmological simulations of varying resolutions to study high-redshift galaxies, the simulations employed in this work are quite unique, since in addition to DM, they have been implemented with a tremendous amount of baryonic physics, including non-equilibrium atomic and molecular chemistry, cooling of gas between $\simeq 10$ and $10^8$ K, star formation, feedback mechanisms, and the full stellar evolution and metal spreading according to the proper yields (for He, C, O, Si, Fe, S, Mg, etc.) and lifetimes in both the PopIII and PopII-I regimes.

\subsection{Identifying LBGs}
\label{identifying_lbgs}
To identify the simulated galaxies that could be observed as LBGs, we start by computing the UV 
luminosity (or absolute magnitude) of each galaxy in the simulation snapshots at $z \simeq 6-12$, 
roughly spaced as $\Delta z \simeq 1$. We consider each star particle to form in a burst, after which it evolves passively. 
As explained above, the simulation used in this work has a detailed treatment of chemodynamics, and consequently we 
are able to follow the transition from metal-free PopIII to metal-enriched PopII/I star formation. If, at the time of 
formation of a star particle, the metallicity of its parent gas particle is less than the critical value of 
${\rm Z_{crit}} = 10^{-4}\Zsun$, we use the PopIII SED, taking into account the mass of the star particle so formed, 
and assuming that the emission properties of PopIII stars remain constant during their relatively short 
lifetime of about $2.5 \times 10^6 {\rm yr}$ \citep{schaerer2002}. On the other hand, if a star particle 
forms out of a metal-enriched gas particle (${\rm Z>Z_{crit}}$), the SED is computed via the population 
synthesis code {\tt STARBURST99} \citep{leitherer1999}, using its mass ($M_*$), stellar metallicity ($Z_*$) and 
age ($t_*$); the age is computed as the time difference between the redshift of the snapshot and the 
redshift of formation of the star particle. Although the luminosity from PopIII stars has been 
included in our calculations, it is negligible compared to that contributed by PopII/I stars 
\citep[see Fig. 9,][]{salvaterra2011}. The composite spectrum for each galaxy is then calculated by summing the SEDs 
of all its (metal-free and metal-enriched) star particles. The intrinsic continuum luminosity, $L_c^{int}$, 
is calculated at rest-frame wavelengths $\lambda = 1350, 1500,1700$ \AA\, at $z \simeq 6,7,8$ respectively; 
though these wavelengths have been chosen for consistency with $J$-band observations, using slightly different 
values would not affect the results in any significant way, given the flatness of the intrinsic spectrum in this 
relatively short wavelength range.

\begin{figure*} 
   \center{\includegraphics[scale=1.01]{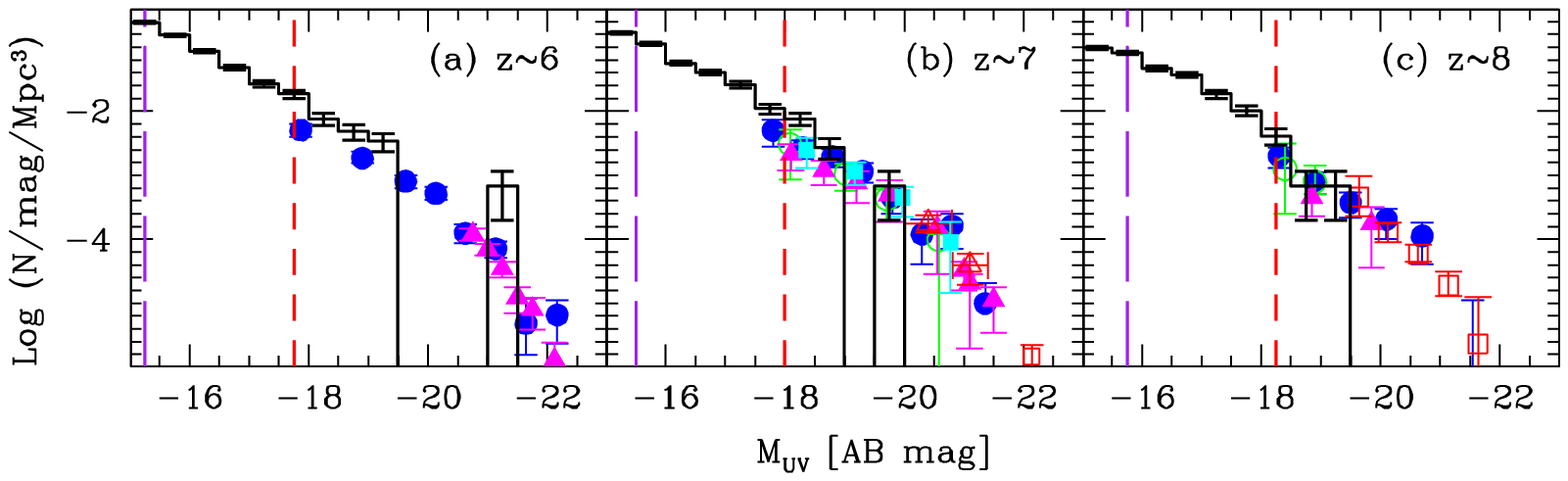}} 
  \caption{The UV LFs of galaxies at $z \simeq 6, 7, 8$ as marked in each panel. 
In all panels, histograms show our theoretical results for galaxies with $M_{UV}<-15$, 
with error bars showing the associated poissonian errors, and points show the observed data. 
The observed LBG UV LFs have been taken from: (a) $z\simeq 6$: \citet[filled circles]{bouwens2007} 
and \citet[filled triangles]{mclure2009}; (b) $z \simeq 7$: \citet[filled squares]{oesch2010}, 
\citet[empty circles]{bouwens2010}, \citet[filled circles]{bouwens2011}, \citet[empty triangles]{castellano2010a}, 
\citet[filled triangles]{mclure2010} and \citet[empty squares]{bowler2012};  
(c) $z \simeq 8$: \citet[empty circles]{bouwens2010}, \citet[filled circles]{bouwens2011}, 
\citet[filled triangles]{mclure2010} and \citet[empty squares]{bradley2012}. The vertical short (long) dashed 
lines in each panel show approximate effective detection limits for HST (JWST) near-infrared imaging.}
\label{fig_lbg_uvlf} 
\end{figure*}

Dust is expected to have only a minor impact on the faint end of the UV LF at 
$z \gsim 6$ where $\langle E(B-V) \rangle<0.05$, as a result of the low stellar mass, age and metallicity of these 
early galaxies \citep[see also][]{salvaterra2011}. Nevertheless,  we still self-consistently compute the dust mass 
and attenuation for each galaxy in the simulation box; we note that while the metal enrichment and mixing have 
been calculated within the simulation as explained in Sec. \ref{sim_des}, the dust masses and attenuation 
are calculated by post-processing the simulation outputs. Dust is produced both by SN and evolved stars in a galaxy. 
However, the contribution of AGB stars becomes progressively less important towards very high redshifts, because the 
typical evolutionary time-scale of these stars ($\geq 1$ Gyr) becomes longer than the age of the Universe 
above $z \gsim 5.7$ \citep{todini-ferrara2001,dwek2007,valiante2009}. We therefore make the hypothesis 
that the dust present in galaxies at $z \ge 6$ is produced solely by SNII. The total dust mass present in each 
galaxy is then computed assuming: (i) ${0.4\, \rm M_\odot}$ of dust is produced per SNII 
\citep{todini-ferrara2001,nozawa2007,bianchi-schneider2007}, (ii) SNII destroy dust in forward shocks with an 
efficiency of about 20\% \citep{mckee1989,seab-shull1983}, (iii) a homogeneous mixture of gas and dust is 
assimilated into further star formation (astration), and (iv) a homogeneous mixture of gas and dust is lost in SNII powered outflows. 

To transform the total dust mass into an optical depth to UV continuum photons, we assume the dust to be made 
up of carbonaceous grains and spatially distributed in a screen of radius $r_d = r_g = 4.5 \lambda r_{200}$ 
\citep{ferrara2000}, where $r_g$ is the radius of the gas distribution, $\lambda = 0.05$ is the average spin parameter for the corresponding galaxy population studied, and $r_{200}$ is the virial radius, assuming that the collapsed region has an over-density which is 200 times the 
critical density at the redshift considered. The reader is referred to \citet{dayal2010a} for complete details of 
this calculation. The observed UV luminosity can then be expressed as $L_c^{obs} = L_c^{int} \times f_c$, 
where $f_c$ is the fraction of continuum photons that escape the galaxy, unattenuated by dust; UV photons 
($\lambda \gsim 1500$\AA) are unaffected by the IGM, and all the UV photons that escape a galaxy unattenuated by dust reach the observer.

\section{Predicted observables for LBGs}
\label{observables_lbg}
Once the UV luminosity of each galaxy in the simulation snapshots at $z \simeq 6-12$ has been calculated as explained above, 
we can study the galaxies that would be detectable as LBGs. Although the current near-infrared observable limit of the deepest
HST surveys corresponds to $M_{UV} \simeq (-17.75,-18.0,-18.25)$ at $z\simeq (6,7,8)$, as a result of the simulation 
resolution we are able to study LBGs that are an order of magnitude fainter ($M_{UV} < -15$); these faint galaxies 
will be detectable with future facilities such as the JWST. As a first test of the simulation, we now compare the UV LFs, 
specific star-formation rate ($sSFR$) and the stellar mass density ($SMD$) of the simulated LBGs to the observed values.

\subsection{UV Luminosity Functions}
\label{lbg_uvlf}
We start by building the UV LFs for all galaxies with $M_{UV} < -15$ in the simulation boxes at $z \simeq 6,7$ and 8. 
It is clear that the luminosity range sampled by the observations and the theoretical model overlap only in the 
range $M_{UV} \simeq -17.75$ to $-19.5$, as shown in Fig. \ref{fig_lbg_uvlf}. This is because the data are limited at the faint end by 
the deepest HST+WFC3/IR data obtained prior to the upcoming UDF12 campaign (HST GO 12498, PI: Ellis), while the models
are limited at the bright end by the size of the simulated volume needed to achieve the mass 
resolution ($\simeq 10^5 \Msun$ in baryons) required to model the PopIII to II transition in high-$z$ galaxies.
Nevertheless, in the magnitude range of overlap, both the amplitude and the slope of the observed and simulated 
LFs are in excellent agreement at all three redshifts. This is a notable success of the model, given that once the SEDs 
and UV dust attenuation have been calculated for each galaxy in the simulated box, except for the caveats involved in running the simulation itself (see Sec. \ref{sim_des}), we have no more free-parameters left to vary when computing the predicted UV LF. 

The simulated UV LFs reproduce well the observed progressive shift of the galaxy population 
towards fainter luminosities/magnitude ($M_{UV}^*$) and/or 
number densities ($\phi^*$) with increasing redshift from $z \simeq 6$ to $8$. Observationally the physical 
driver of this evolution is not yet clear, with the Schechter function fits of some authors favouring 
luminosity evolution \citep{mannucci2007, castellano2010b, bunker2010, bouwens2011}, while the results of 
other studies appear to favour primarily density evolution \citep{ouchi2009, mclure2010}. However,
by tracing the histories of the individual galaxies in the our simulation we can hope to shed 
light on the main physical drivers of this population evolution (see Sec. 5).

\begin{figure} 
   \center{\includegraphics[scale=0.45]{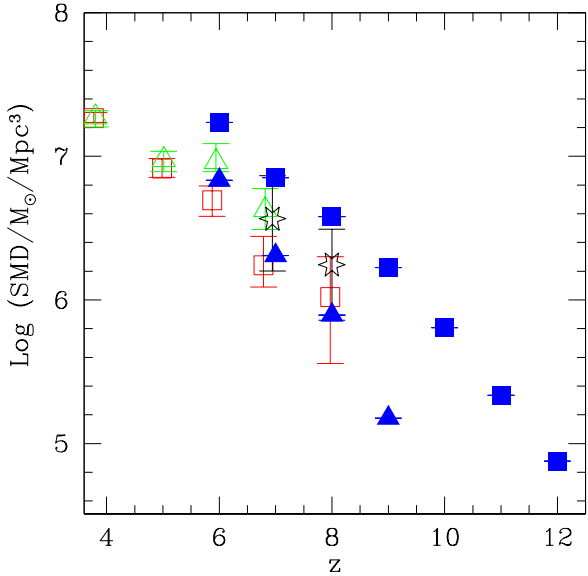}} 
  \caption{Stellar mass density ($SMD$) as a function of redshift. 
Filled squares (triangles) show our theoretical predictions for galaxies with 
$M_{UV}<-15$ ($M_{UV}<-18$) at each redshift, with error bars showing the 
associated poissonian errors. Empty points show the magnitude-limited $SMD$ values 
inferred observationally for galaxies with $M_{UV}<-18$ by  \citet[empty stars]{labbe2010a,labbe2010b} 
and \citet[empty triangles]{gonzalez2011}; empty squares show the $SMD$ 
inferred by \citet{stark2012} for the same magnitude limit after correcting the 
stellar masses for nebular emission-line contributions to the broad-band fluxes
(assuming that the nebular line rest-frame equivalent width evolves with redshift).}
\label{smd_fig} 
\end{figure}

Our simulations also reproduce well the observed steep faint-end slope of the UV LF, consistent 
with $\alpha \simeq -2$  at all three redshifts. There is, however, still considerable debate 
over the precise value of $\alpha$ observed at these redshifts \citep{oesch2010,mclure2010,bouwens2011,bradley2012}. 
The faint-end slope of the LF has important implications for reionization since these faint galaxies 
are expected to be the major sources of \HI ionizing photons 
\citep[see e.g.][]{choudhury-ferrara2007,robertson2010,salvaterra2011,ferrara2012}. The current uncertainty in the 
value of $\alpha$ should be clarified somewhat by UDF12 (Mclure et al., in preparation), 
and should be resolved by JWST with its forecast ability to 
reach absolute magnitude limits of $M_{UV} \simeq (-15.25, -15.5, -15.75)$ at $z \simeq (6,7,8)$ respectively.

\subsection{Stellar mass density}
\label{smd}
The stellar mass, $M_*$, is one of the most fundamental properties of a high-$z$ galaxy, encapsulating information about its 
entire star-formation history. However, achieving accurate estimates 
of $M_*$ from broad-band data is difficult because it depends on the assumptions made regarding the 
IMF, SF history, strength of nebular emission, age, stellar metallicity and dust; the latter three parameters 
are degenerate, adding to the complexity of the problem. Although properly constraining 
$M_*$ ideally requires rest-frame near infra-red data which will be provided by future instruments 
such as MIRI on the JWST, broad-band HST+{\it Spitzer} data have already been used to infer the 
contribution of galaxies brighter than $M_{UV} = -18$ to the growth in total stellar mass 
density (SMD = stellar mass per unit volume) with decreasing redshift 
\citep{stark2009,labbe2010a,labbe2010b,gonzalez2011}. Encouragingly, as shown in Fig. \ref{smd_fig}, 
the observed growth in $SMD$ is reproduced very well by integrating the stellar masses 
of the simulated galaxies brighter than $M_{UV}<-18$; the theoretically calculated SMD of these galaxies drops to zero at $z > 9$ since there are no galaxies massive enough to be visible with this magnitude cut in the volume simulated. Further, as a result of their much larger numbers, galaxies with $-18 \leq M_{UV} \leq -15$ contain about {\it 1.5 times} the mass as compared to the larger and more luminous galaxies that have been observed as of date; the JWST will be instrumental in shedding light on the properties of such faint galaxies, in which most of the stellar mass is locked up at these high-$z$.

\begin{figure} 
   \center{\includegraphics[scale=0.45]{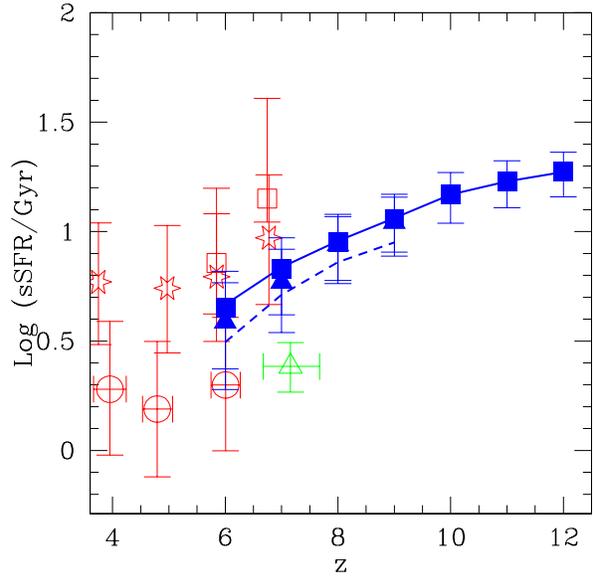}} 
  \caption{Specific star-formation rate ($sSFR$) as a function of redshift. Filled squares (triangles) 
show the theoretical results for galaxies with $M_{UV}<-15$ ($M_{UV}<-18$) 
at each redshift, with error bars showing the associated poissonian errors. 
Empty points show the $sSFR$ values inferred observationally by 
\citet[empty circles]{stark2009}, 
\citet[empty triangles]{gonzalez2010}; empty stars and squares show the $sSFR$ 
inferred by \citet{stark2012}, after correcting the stellar masses for nebular emission lines 
assuming that the nebular line rest-frame equivalent width at $z \simeq 4-7$ is 
the same as that as $z\simeq 3.8-5$, and alternatively that the nebular line rest-frame equivalent width 
evolves with redshift, respectively. The solid and dashed lines show the evolution of $sSFR$ as inferred from the simulation for galaxies 
with $M_*=10^{6-8}\Msun$ and $M_*>10^8 \Msun$, respectively.}
\label{ssfr_fig} 
\end{figure}

\begin{table*} 
\begin{center} 
\caption {For the redshift, $z$, given in Column 1, we show the following ratios: 
$\Sigma M_{*,MB}(z)/\Sigma M_{*,z=6}$ (column 2), $\Sigma M_{*,MB}(z)/\Sigma M_{*,allprog}(z)$ (column 3), 
$\Sigma M_{*,allprog}(z)/\Sigma M_{*,z=6}$ (column 4), $\Sigma M_{sf}(z,z-1)/\Sigma M_{*,z=6}$ (column 5), 
$\Sigma M_{acc}(z,z-1)/\Sigma M_{*,z=6}$ (column 6) and $\Sigma M_{acc}(z,z-1)/\Sigma M_{sf}(z,z-1)$ (column 7). 
Here, $\Sigma M_{*,z=6}$ represents the total stellar mass at $z\simeq 6$, summed over all $z \simeq 6$ LBGs 
in the simulation (and has a value $\Sigma M_{*,z=6} = 5.02 \times 10^{10} \Msun$), $\Sigma M_{*,MB}(z)$ 
is the stellar mass summed over all the major branch progenitors of $z \simeq 6$ LBGs at the given redshift, 
$\Sigma M_{*,allprog}(z)$ is the total stellar mass in all the other progenitors of $z \simeq 6$ LBGs 
(i.e. excluding the major branch one) at the redshift $z$, and $\Sigma M_{sf}(z,z-1)$ and $\Sigma M_{acc}(z,z-1)$ 
represent the total stellar mass assembled by the major branch progenitors of $z \simeq 6$ 
LBGs by star formation and mergers respectively, between the redshifts $z$ and $z-1$ 
(see Sec. \ref{sf_acc} for details).} 
\begin{tabular}{|c|c|c|c|c|c|c|c}
\hline 
$z$& $\frac{\Sigma M_{*,MB}(z)}{\Sigma M_{*,z=6}}$ & $\frac{\Sigma M_{*,MB}(z)}{\Sigma M_{*,allprog}(z)}$ & $\frac{\Sigma M_{*,allprog}(z)}{\Sigma M_{*,z=6}}$ & $\frac{\Sigma M_{sf}(z,z-1)}{\Sigma M_{*,z=6}}$ & $\frac{\Sigma M_{acc}(z,z-1)}{\Sigma M_{*,z=6}}$ & $\frac{\Sigma M_{acc}(z,z-1)} {\Sigma M_{sf}(z,z-1)}$ \\  
\hline 
$7$& $0.37$ & $6.1$ & $6.1 \times 10^{-2}$  & $0.56$ & $6.1 \times 10^{-2}$ & $0.109$\\
$8$& $0.19$ & $3.7$ & $5.1 \times 10^{-2}$ &  $0.16$ & $1.5 \times 10^{-2}$ & $9.3 \times 10^{-2}$ \\
$9$& $7.9 \times 10^{-2}$ & $2.4$ & $3.2 \times 10^{-2}$ &  $0.10$ & $1.15 \times 10^{-2}$ & $1.1 \times 10^{-2}$ \\
$10$& $2.9 \times 10^{-2}$ & $2.0$ & $1.4 \times 10^{-2}$ & $4.7 \times 10^{-2}$ & $2.55 \times 10^{-3}$ & $5.4 \times 10^{-2}$\\
$11$& $1.0 \times 10^{-2}$ & $1.8$ & $5.5 \times 10^{-3}$ & $1.8 \times 10^{-2}$ & $8.02 \times 10^{-4}$ & $4.3 \times 10^{-2}$\\
$12$& $3.4 \times 10^{-3}$ & $1.8$ & $2.1 \times 10^{-3}$  & $6.6 \times 10^{-3}$ & $8.44 \times 10^{-5}$ & $1.2 \times 10^{-2}$\\
\hline
\label{table1} 
\end{tabular} 
\end{center}
\end{table*}

\subsection{Specific star formation rates}
\label{ssfr}
The specific star-formation rate ($sSFR$) is an important physical quantity 
that compares the current level of SF to the previous SF history of a galaxy.
It also has the advantage of being relatively unaffected by the assumed IMF.
There is now a considerable body of observational evidence indicating that 
the typical $sSFR$ of star-forming galaxies rises by a factor of about 40 
between $z = 0$ and $z \simeq 2$ \citep[e.g.][]{daddi2007}, but then 
settles to a value consistent with $2-3 \, {\rm Gyr^{-1}}$ at $z \simeq 3-8$ 
\citep{stark2009,gonzalez2010,mclure2011,stark2012,labbe2012}. 

This constancy of $sSFR$ at high redshift has proved somewhat unexpected and 
difficult to understand, given that theoretical models predict that $sSFR$ 
should trace the baryonic infall rate which scales as $(1+z)^{2.25}$ \citep{neistein2008,weinmann2011}; 
according to this calculation, the typical $sSFR$ should increase by a factor of about 9 
over the redshift range $z \simeq 2 - 7$. 
However, recently, \citet{stark2012} and \citet{labbe2012} have suggested that 
at least part of this discrepancy might be due to a past failure to 
account fully for the (potentially growing) level of nebular emission-line contamination of 
broad-band photometry at high-$z$, as suggested by \citet{schaerer2009}; 
indeed, \citet{stark2012} estimate that the emission-line corrected $sSFR$ may rise by a factor of about 5 between $z \simeq 2
-3$ and $z \simeq 7$.

From the simulation snapshots, we find that typical $sSFR$ decreases with increasing $M_*$. 
However, this mass dependence is relatively modest, with the $sSFR$ of galaxies with 
$M_* = 10^{6-8} {\rm M_\odot}$ being only $\simeq  1.5$ times larger 
than the $sSFR$ for galaxies with $M_* > 10^{8} {\rm M_\odot}$ over the redshift 
range $z \simeq 6-9$ as shown in Fig. \ref{ssfr_fig} \citep[see also Sec. 3.4,][]{dayal2012} (note that 
galaxies with $M_*>10^8 \Msun$ have not had time to evolve before $z \simeq 9$ in the simulated volume).
Examples of the evolution of $sSFR$ for individual simulated galaxies of varying stellar mass 
are shown in Fig. \ref{ssfr_fnz} in the appendix. 

More dramatic is the predicted trend with redshift above $z = 6$. 
As shown in Fig. \ref{ssfr_fig}, averaged over all LBGs with $M_{UV}<-15$,
the typical $sSFR$ inferred from our simulations rises by a factor $\simeq 4$ from $z \simeq 6$ to $z \simeq 12$ 
(from $sSFR = \simeq 4.5\,{\rm Gyr^{-1}}$ to $sSFR \simeq 18.6\, {\rm Gyr^{-1}}$).
This is a strong prediction for future observations, and it is heartening to note that,
in the redshift range probed by current data ($z \simeq 6-7$), our predicted 
$sSFR$ values are in excellent agreement with the most recent observational 
results, the range of which is dominated by the degree of nebular emission-line correction applied.

\section{Stellar mass assembly}
\label{mass_assembly}
Having shown that the simulation successfully reproduces the key current observations of the galaxy population
at $z = 6 - 8$, we can legitimately proceed to use 
our model to explore how the galaxies on the UV LF shown in 
Fig. \ref{fig_lbg_uvlf} build up their stellar mass. We  again use the galaxies at $z \simeq 6$ 
as our reference point, and in what follows 
all galaxies with $M_{UV} < -15$ at $z \simeq 6$ (which number 871) are 
colloquially referred to as $z \simeq 6$ LBGs and their stellar mass is denoted as 
$M_{*,z=6}$. 

\begin{figure*} 
   \center{\includegraphics[scale=1.0]{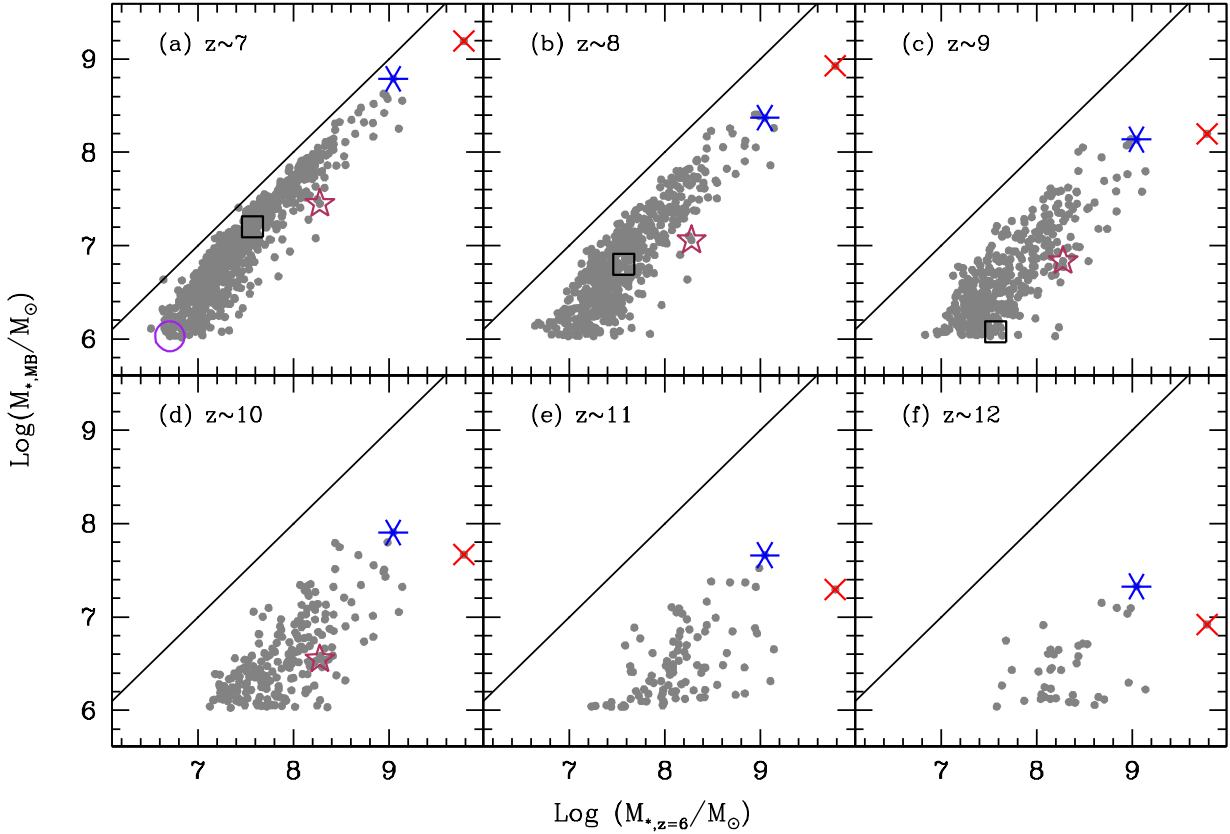}} 
  \caption{The stellar mass of the major branch progenitor as a function 
of the $z \simeq 6$ stellar mass, for $z\simeq6$ LBGs (small points) at the redshifts marked in each panel: 
$z\simeq 7-12$. In each panel, the large symbols show the results for the 5 different example $z \simeq 6$ 
LBGs discussed in the text: galaxy A (empty circle), B (empty square), C (empty star), D (asterix) and E (cross).
The solid line indicates $M_{*,MB}=M_{*,z=6}$ to guide the eye.}
\label{msmb} 
\end{figure*}

Before discussing how $z \simeq 6$ LBGs assemble their stellar mass, we briefly digress to explain 
how we identify their progenitors and the major branch of their merger trees between $z \simeq 7$ and $z \simeq 12$, 
in snapshots spaced by $\Delta z\simeq 1$. We start our analysis from the simulation snapshot at $z \simeq 7$. 
Since each star particle is associated with a bound structure, in step one, by matching the star 
particles in the snapshot at $z \simeq 7$ and $z \simeq 6$, we identify all the progenitors of each 
$z \simeq 6$ LBG at $z \simeq 7$; alternatively, each $z \simeq 7$ galaxy has a $z \simeq 6$ ``descendant" 
galaxy associated with it. For each $z \simeq 6$ LBG, the $z \simeq 7$ progenitor with the largest total 
(DM+gas+stellar) mass is then identified as the major branch of its merger tree at $z \simeq 7$. 
In step two, matching the star particles in the simulation snapshots at $z \simeq 7$ and $z \simeq 8$, 
each $z\simeq 8$ galaxy is associated to a ``descendant" galaxy at $z \simeq 7$  
(i.e. we find all the progenitors of each $z \simeq 7$ galaxy). 

Again, for each $z \simeq 7$ galaxy, the $z \simeq 8$ progenitor with the largest total mass 
is identified as the major branch of its merger tree at $z \simeq 8$. Since we know the 
IDs of the major branch  and all the progenitors of each $z \simeq 6$ LBG at $z \simeq 7$ 
from step one above, and the ID of the $z \simeq 8$ major branch and all progenitors of each $z \simeq 7$ 
galaxy from step two, for each $z \simeq 6$ LBG we can identify (a) all the progenitors at $z \simeq 
7,8$, and (b) the major branch of the merger tree at $z \simeq 7,8$. 
This same procedure is then followed to find all the progenitors, and the major branch of the merger 
tree for each $z \simeq 6$ LBG, at $z \simeq 9,10,11,12$; we end the calculations at 
$z \simeq 12$ because only a few thousand star particles 
(and a few hundred bound structures) exist at this epoch in the simulated volume, 
leading to very poor number statistics. 

\subsection{Assembling the major branch}
\label{mb_assembly}
We now return to our discussion on the stellar mass assembly of the major branch, 
$M_{*,MB}$, for $z \simeq 6$ LBGs. As expected from the hierarchical 
structure formation scenario where larger systems build up from the merger of smaller ones, 
the earlier a system starts assembling, the larger its final mass is likely to be; 
alternatively, this implies that the progenitors of the largest systems start assembling 
first, with the progenitors of smaller systems assembling later. A 
natural consequence of this behaviour is that it leads to a correlation 
between $M_{*,z=6}$ and $M_{*,MB}$, i.e. at any given $z$, the major branch 
stellar mass is generally expected to scale with the final stellar mass at $z \simeq 6$. 

\begin{figure*} 
   \center{\includegraphics[scale=1.0]{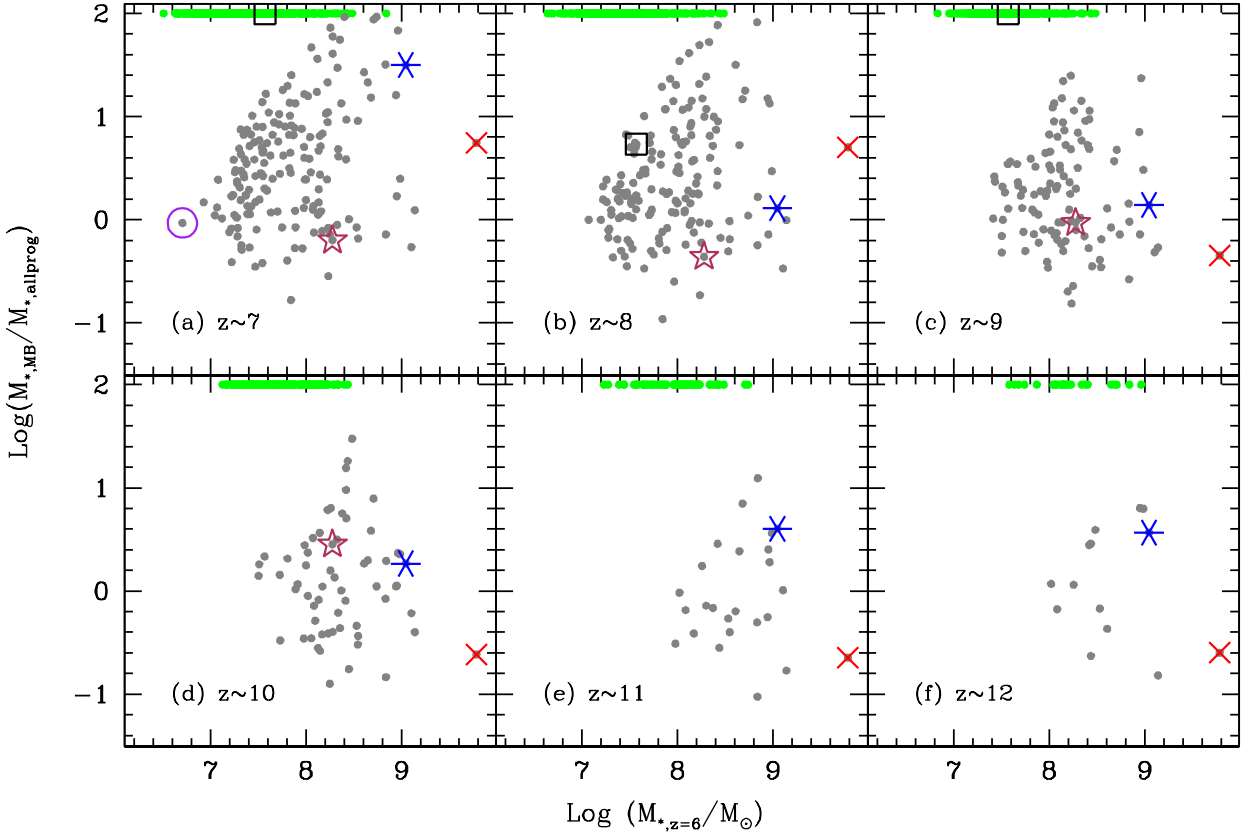}} 
  \caption{The ratio of the stellar mass of the major branch progenitor and all the other 
progenitors (i.e. excluding the major branch progenitor) for each $z\simeq6$ LBG, as a function 
of the $z \simeq 6$ stellar mass. The panels show the results for different redshifts, 
$z\simeq 7-12$, and the small grey (green) points show the results for LBGs with more than 
one (a single) progenitor. Since galaxies with a single progenitor have an undefined 
value of $M_{*,MB}/M_{*,allprog}$, we use an arbitrary ratio of $10^2$ to show these galaxies. 
The large symbols show the results for the 5 different example  $z \simeq 6$ LBGs discussed in the text: 
A (empty circle), B (empty square), C (empty star), D (asterix) and E (cross).}
\label{rat_mball} 
\end{figure*}

This behaviour can be seen clearly in Fig. \ref{msmb}: firstly, we see that the 
progenitors of the most massive $z \simeq 6$ LBGs start assembling first, 
and progenitors of increasingly smaller systems appear with decreasing $z$. 
Indeed, while the progenitors of $z \simeq 6$ LBGs with $M_* \gsim 10^8 \Msun$ 
already exist at $z \simeq 12$, there is a dearth of progenitors of the lowest-mass 
galaxies, with $M_* \simeq 10^{7} \Msun$, which start building up as late as $z \simeq 9$. In other words, the number of points in Fig. \ref{msmb} (and Figs. \ref{rat_mball}, \ref{rat_sfmer} and \ref{muvmb} that follow) decrease with increasing redshift since only the major branches of the most massive $z \simeq 6$ LBGs can be traced all the way back to $z \simeq 12$. It can also be seen from the same figure that there is a broad trend for 
the $z\simeq 6$ LBGs with the largest $M_{*,z=6}$ values to have largest values of $M_{*,MB}$ 
at all the redshifts studied $z \simeq 7-12$, although the scatter grows substantially,
and only a few hundred progenitors exist at $z \geq 11$. The wide spread in $M_{*,MB}$ 
for a given final $M_{*,z=6}$ value, points to the varied stellar mass build-up histories of
the galaxies in the simulation; the relation between $M_{*,MB}$ and $M_{*,z=6}$ inevitably tightens 
with decreasing redshift (see also Table \ref{table1}). 

Finally, we note that, averaged over all $z \simeq 6$ LBGs, only about $0.3\%$ of the final stellar mass 
has been built up by $z \simeq 12$. The major branch of the merger tree steadily builds up 
in mass at a rate (mass gained/Myr) that increases with decreasing redshift such that 
about $(1,3,8,19,37)$\% of the final stellar mass is built up by 
$z \simeq (11,10,9,8,7)$ as shown in Table 1; this implies that $z \simeq 6$ LBGs gain the bulk of 
their stellar mass ($\approx 63 \%$) in the $\simeq 150$\,Myr between $z \simeq 7$ and $z \simeq 6$, 
and only about a third is assembled in the preceding $\simeq 400$\,Myr between $z \simeq 12$ and $z \simeq 7$. 
This rapidly increasing stellar mass growth rate with decreasing $z$ probably results from 
negative mechanical feedback becoming less important as galaxies grow more massive: 
even a small amount of star formation in the tiny potential wells of early progenitors 
can lead to a partial blowout/full blow-away of the interstellar medium (ISM) gas, 
suppressing further star formation (at least temporarily). These progenitors must then wait 
for enough gas to be accreted, either from the IGM or through mergers, 
to re-ignite star formation; negative feedback is weaker in more massive 
systems due to their larger potential wells \citep[see also][]{maio2011}. We note that in the simulation used in this work, galactic winds have a fixed velocity of $500 \, {\rm km \, s^{-1}}$ with a mass upload rate equal to twice the local star formation rate, and carry away a fixed fraction ($25\%$) of the SN energy (for which the canonical value of $10^{51} \,{\rm ergs}$ is adopted). Increasing the mass upload rate (fraction of SN energy carried away by winds) would lead to a decrease (increase) in the wind velocity as it leaves the galactic disk, making mechanical feedback less (more) efficient. 

To help illustrate and clarify the above points, we show the stellar mass growth of five 
different $z \simeq6$ LBGs with $\log [M_{*,z=6}/\Msun] = (6.7,7.6,8.3,9.1,9.8)$; 
these are labelled (A,B,C,D,E) respectively. As seen from panel (f) of Fig. \ref{msmb}, 
the progenitors of the two most massive example galaxies, D and E, appear as early 
as $z \simeq 12$, making some of their stellar content at least as old as 550\,Myr by $z \simeq 6$. 
However, although these galaxies started building up their stellar mass early, their stellar 
mass-weighted ages are, of course, substantially younger, at $(227,132)\, {\rm Myr}$ respectively; 
consistent with the statement made above, the bulk of their stellar mass was formed in the 
150-200\,Myr immediately prior to $z \simeq 6$ and these galaxies have only assembled 
about $(1.9,0.1)\%$ of their final stellar mass at $z \simeq 12$. The value of $M_{*,MB}$ 
steadily builds up, either by merging with the other progenitors of the ``descendant" galaxy, 
or due to star formation in the major branch itself such that, by $z \simeq 10$, these galaxies 
have increased in mass by a factor of about 4 (6). At $z \simeq 10$ (9), the progenitor of galaxy C (B) appears, 
putting the age of the oldest stars in this galaxy at $>$ 450 (380)\,Myr by $z \simeq 6$; again the bulk of the 
stellar mass for both these galaxies forms in the $\simeq 156$\,Myr between $z \simeq 7$ and $z \simeq 6$. 
The progenitor of galaxy A finally appears at $z \simeq 7$, having formed a tiny stellar mass of 
$M_{*,MB} \simeq 10^{6}\Msun$. Although by $z \simeq 7$, the five galaxies illustrated have individually built up 
between 15-50\% of their final stellar mass, their average $z \simeq 7$ stellar mass is about 
30\% of the final value; this is very consistent with the global average value of 37\%.

Finally, we remind the reader that for our calculations, we have used galaxies that fulfil three criterion: $M_h \geq 10^{8.5} \, \Msun$, more than $4N$ gas particles and a minimum of 10 star particles. Increasing the minimum number of star particles required will naturally lead of a number of galaxies no longer being considered `resolved'. For example, galaxy A which has a total of 48 star particles at $z \simeq 6$ would not have been identified as a galaxy imposing a cut of 50 minimum star particles. However, we also note that increasing the minimum number of star particles from 10 to 50 only affects the faintest galaxies between $M_{UV}=-15 $ and $-15.5$ for $z \simeq 7-12$ as shown in Fig. \ref{res_uvlf}.

\subsection{Relative growth: major branch versus all progenitors}
\label{rel_mb_all}
\begin{figure*}
\begin{minipage}{0.5\linewidth}
\center{\includegraphics[scale=0.45]{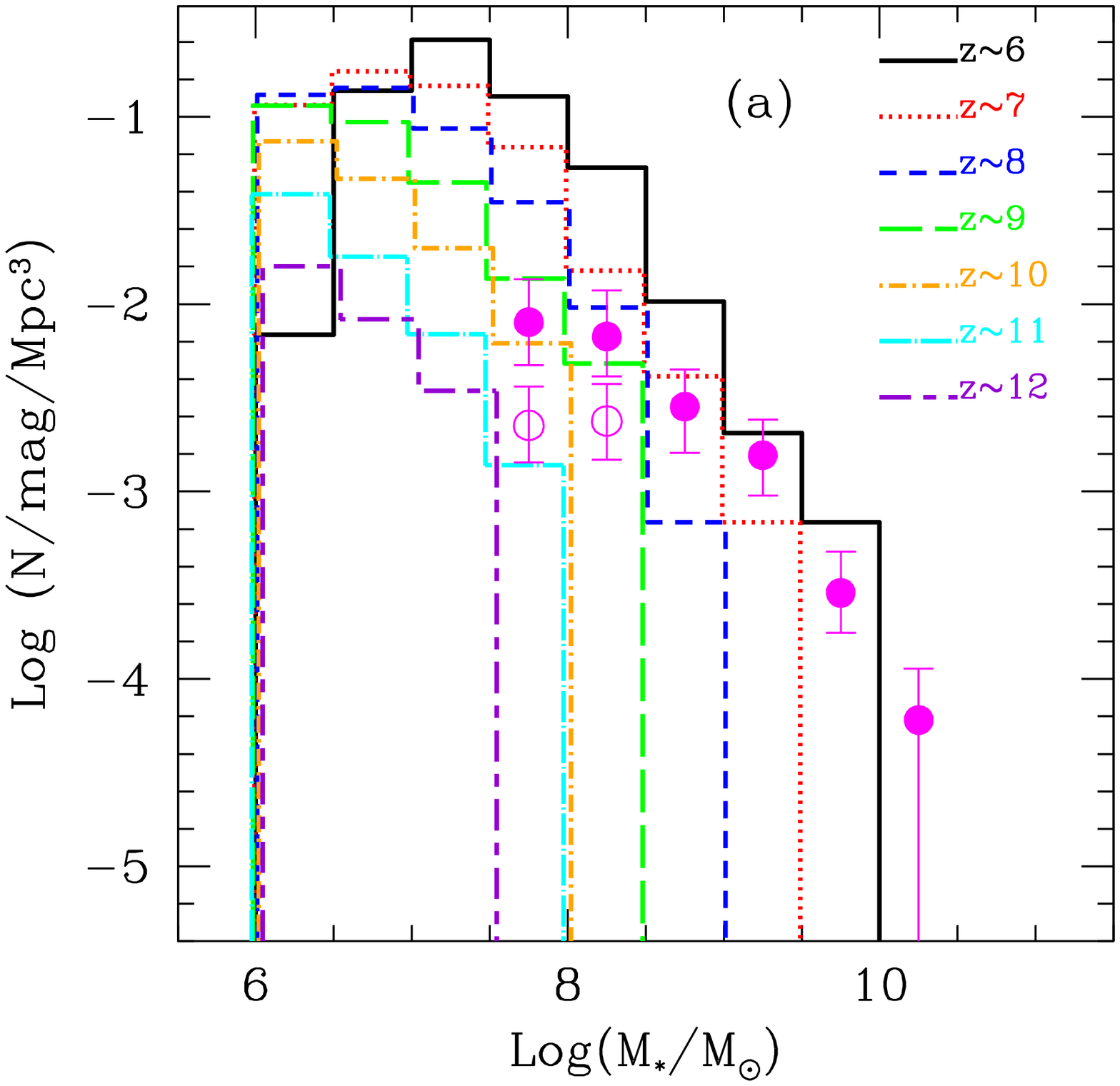}}
\end{minipage}%
\begin{minipage}{0.5\linewidth}
\center{\includegraphics[scale=0.45]{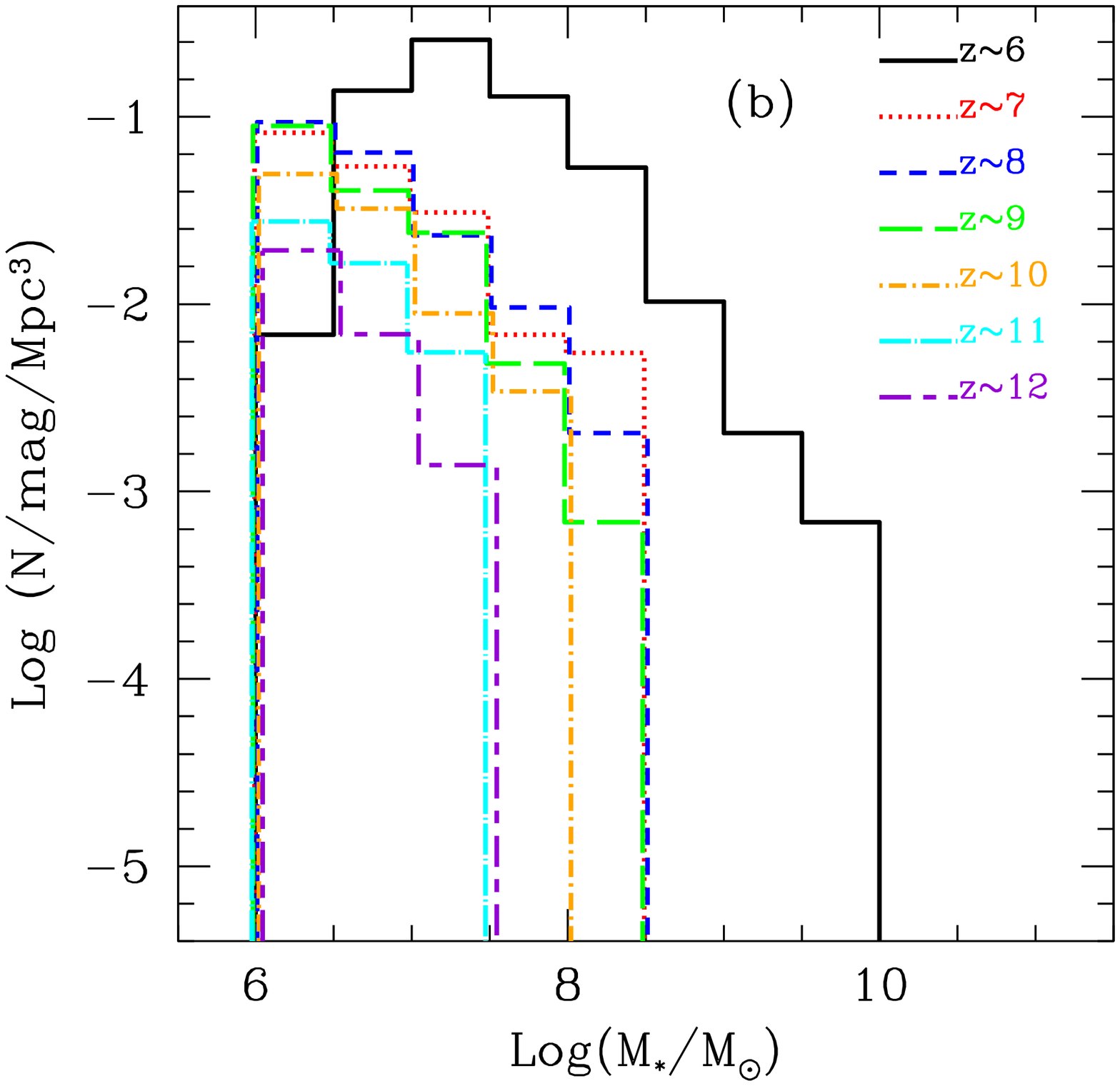}}
\end{minipage}
\label{massfn}
\caption{The stellar mass functions for the progenitors of $z \simeq 6$ LBGs: the left and right-hand panels 
show the results for the major branch of the merger tree, and all the other progenitors 
(i.e. excluding the major branch), respectively. In each panel, the lines show the 
results at different redshifts: $z\simeq 7$ (dotted red), $z\simeq 8$ (short dashed blue), $z\simeq 9$ (long dashed green), 
$z\simeq 10$ (dot-short dashed orange), $z\simeq 11$ (dot-long dashed cyan), $z\simeq 12$ 
(short dashed-long dashed violet). For comparison, the solid line in each panel shows the 
stellar mass function for $z \simeq 6$ LBGs. 
In the left-hand panel, filled (empty) points show the corrected (uncorrected) $z \simeq 6$ 
stellar mass functions inferred observationally by \citet{gonzalez2011}. In both panels, we have slightly displaced the various mass functions horizontally for clearer visualization.}
\label{fig_massfn}
\end{figure*}

We now ask how the growth of $M_{*,MB}$ compares to the stellar mass growth 
of all the {\it other} progenitors (i.e. excluding the major branch itself) of $z\simeq 6$ 
LBGs, $M_{*,allprog}$. Naively it might be expected that when the major branch of a 
galaxy first forms, its stellar mass might be low compared to that contained in all the 
other progenitors; the ratio $M_{*,MB}/M_{*,allprog}$ would increase with time 
as $M_{*,MB}$ builds up, becoming the dominant stellar mass repository. This is indeed the general behaviour found for the progenitors of $z \simeq 6$ LBGs, where the value of the total stellar mass contained in all the major branch 
progenitors of $z \simeq 6$ LBGs, $\Sigma M_{*,MB}$, increases from $1.8$ to $ 6.1$ 
times the total mass contained in all the other progenitors, $\Sigma M_{*,allprog}$, 
as the redshift decreases from $z \simeq 12$ to $z \simeq 7$ (as shown in Fig. \ref{rat_mball}; 
see also Table \ref{table1}); while between $z \simeq 12$ and $z \simeq 10$, all the other 
progenitors put together hold almost half as much stellar mass as the major branch, 
this value rapidly decreases thereafter, such that by $z \simeq 7$, the other progenitors 
together contain only about 15\% of the mass in the major branch. 

The ratio $M_{*,MB}/M_{*,allprog}$ increases very slowly between $z \simeq 12$ and $z \simeq 10$ 
because, at these redshifts, both the major branch and all the other progenitors 
seem to be growing in stellar mass at the same pace. However, at $z \leq 10$, the major 
branch progenitor starts growing more quickly, possibly due to the fact that, 
as the major branch becomes increasingly more massive, it is increasingly 
less affected by the negative mechanical feedback (i.e. SN powered outflows) 
that can quench star formation in the other smaller progenitors.

Further, as discussed above in Sec. \ref{mb_assembly}, the progenitors 
of successively less massive $z \simeq 6$ LBGs appear with decreasing $z$, 
leading to a scatter of more than an order of magnitude in $M_{*,MB}/M_{*,allprog}$, 
at any of the redshifts shown in Fig. \ref{rat_mball} due to their varied assembly histories. 
Indeed, although on average the major branch is the heavyweight in terms of the stellar mass content, 
at any of the redshifts studied, there are always a few galaxies where the major branch 
only contains about 10\% of the total stellar mass at that redshift. 
Finally, we note that, at all redshifts, a number of progenitors show a value of 
$M_{*,MB}=M_{*,allprog}$: these are galaxies with a single progenitor, the major-branch one. 
Although such galaxies do not have a defined $M_{*,MB}/M_{*,allprog}$ value, 
we have used an arbitrary value of $10^2$ to represent them in Fig. \ref{rat_mball}.

The five example LBGs discussed in Sec. \ref{mb_assembly} help to illustrate these points:  
galaxy E, the most massive galaxy in the simulation box, assembles from tens of tiny progenitors, 
and at $z \simeq 12$ its major branch contains only about 25\% of the stellar mass in place at that epoch,  
with the bulk (75\%) contained in all the other progenitors. On the other hand, galaxy D 
has a very different assembly history; at $z \simeq 12$ it already has major branch progenitor 
that is about three times more massive than all the other progenitors put together. 
Between $z \simeq 12$ and $z \simeq 9$ the value of $M_{*,MB}/M_{*,allprog}$ remains 
almost unchanged for E; although $M_{*,MB}$ has doubled for E between these redshifts. 
This is an example of a case where all the other progenitors put together grow in mass as rapidly 
as the major branch. For galaxy C, on the other hand, all the other progenitors grow more quickly 
than the major branch between $z \simeq 10$ and $z \simeq 8$, leading to a decrease 
in the value of $M_{*,MB}/M_{*,allprog}$ with time. Galaxy B is a classic example 
of a galaxy that starts off with a single progenitor (the major branch one) at $z \simeq 9$, and 
then briefly has two progenitors at $z \simeq 8$, before these merge so that the major 
branch again holds all the stellar mass by $z \simeq 7$. At $z \simeq 7$, 
the recently formed major branch of A already holds the majority of the total stellar mass assembled; 
by this redshift, D and E have assembled a major branch that is about 30 and 6 times heavier in stellar mass 
than all the other progenitors together.

\begin{figure*} 
   \center{\includegraphics[scale=1.0]{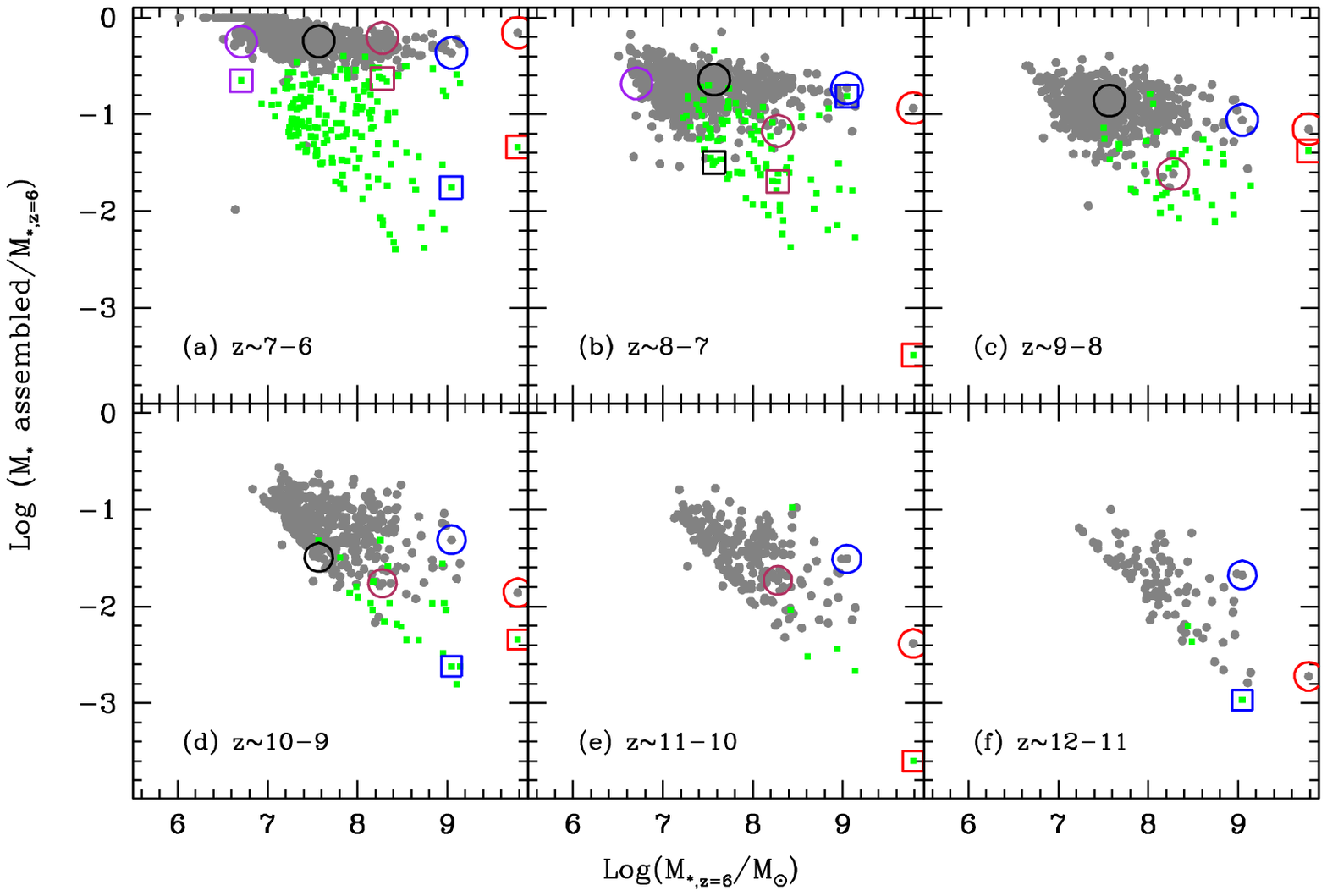}} 
  \caption{The fraction of the total stellar mass assembled in $z\simeq 6$ LBGs by star formation in the major branch (grey circles), and by merging of the minor progenitors
(green squares), in the redshift ranges marked in each panel, as a function of the final stellar mass of each object at $z \simeq 6$. 
The large squares (circles) in each panel show the mass assembled by the 5 example $z \simeq 6$ LBGs by mergers (star formation)
in each redshift interval: A (purple), B (black), C (maroon), D (blue) and E (red). Note that when any galaxy/major branch progenitor 
has only one progenitor at the previous redshift, all the mass is assembled because of star formation; 
the mass brought in by mergers is, by definition, zero. }
\label{rat_sfmer} 
\end{figure*}

We now summarize the relationship between $M_{*,z=6}$, $M_{*,MB}$ and $M_{*,allprog}$ 
by representing these quantities in terms of their mass functions at all the redshifts studied (see Fig. \ref{fig_massfn}). 
We start by noting that, while the theoretical $z \simeq 6$ LBG stellar mass function is in 
broad agreement with the observed one for $M_* > 10^9 \Msun$, it is much steeper than currently 
inferred from the observations for lower values of $M_*$. 
This may be due to the fact that the observational stellar mass function has been constructed 
by inferring $M_*$ from the UV luminosities of LBGS with $M_{UV}<-18$ \citep{gonzalez2011}, 
and is incomplete at the low-mass end. Further, the theoretical stellar mass function 
of $z \simeq 6$ LBGs peaks at $M_* \simeq 10^{7-7.5} \Msun$, as shown in panel (a) 
of Fig. \ref{fig_massfn}. The decrease in the number density of the more massive objects 
is only to be expected from the hierarchical structure formation model. The decrease in the 
number density of lower stellar mass systems, on the other hand, arises from the fact that 
due to their low stellar masses, not many of these galaxies produce enough UV luminosity 
to be visible as LBGs with $M_{UV} < -15$, i.e. this drop is not an artefact of the simulation resolution; the requirement that a galaxy 
has at least 10 star particles corresponds to a resolved stellar mass of $10^6 \Msun$. 

In panel (a) of Fig. \ref{fig_massfn}, we show the evolving stellar mass function of the major branch progenitors at $z \simeq 7-12$ which, as expected, shifts to progressively lower $M_*$ values with increasing redshift. The major branch progenitors completely dominate the high mass end of the  evolving stellar mass function, as becomes clear by comparison with panel (b) which shows the stellar mass functions of all the other progenitors 
(i.e. excluding the major branch) of $z \simeq 6$ LBGs. This shows that 
(i) these are always less massive than $10^{8.5} \Msun$ at any of the redshifts considered, with this mass threshold 
decreasing with increasing $z$, as expected, and (ii) the contribution of the other progenitors is 
largest at the highest $z \simeq 10-12$ where these contain about half the stellar mass in the major branch.
When compared, it is clear that the major branch dominates over the contribution 
from all the other progenitors for $M_* \geq 10^7 \Msun$. 
However, at the low mass end ($M_* \leq 10^7 \Msun$), there is an abundance of tiny progenitors that contributes
to the mass function by a factor of about 1.25 (0.7) compared to the major branch for $z \simeq 12$ ($z \leq 11$).  

\subsection{Mass growth driver: star formation or mergers?}
\label{sf_acc}
We now return to Figs. \ref{msmb} and \ref{rat_mball} to consider an interesting question: 
what is the relative contribution of mergers/accretion ($M_{acc}$) versus star formation 
($M_{sf}$) to building up the total stellar mass of the major branch? A first hint 
of the answer can be obtained by comparing the population average numbers 
as presented in columns 1 and 3 of Table 1  
(see also Secs. \ref{mb_assembly} and \ref{rel_mb_all}) for $z \simeq 6$ LBGs: 
at $z \simeq 7$, the total stellar mass in the major branch has a value $0.37 M_{*,z=6}$, 
and all the other progenitors contain a total stellar mass of $0.06 M_{*,z=6}$. 
This implies that the merger of all the other progenitors into the major branch can increase the mass up 
to $0.43 M_{*,z=6}$, and thus that the remaining 57\% of $M_{*,z=6}$ must be produced by star formation in the 
redshift interval $z \simeq 7$ to $z \simeq 6$. This then implies that, on average, in the redshift 
interval $z \simeq 7 - 6$, stellar mass growth due to star formation dominates 
growth due to the accretion of pre-existing stars in the other progenitors by a factor
$0.57/0.06 \simeq 10$.

For a more complete answer, we have carried out this calculation on a 
galaxy-by-galaxy basis. 
At any redshift, the stellar mass accreted by the major branch due to mergers between 
redshifts $z$ and $z-1$ is calculated as the difference between the mass of {\it all} 
the major branch progenitors and the major branch mass at $z$, i.e. $M_{acc} (z,z-1) = M_{*,allprogMB}(z) - M_{*,MB}(z)$. 
The stellar mass assembled due to star formation is calculated as the difference between the major 
branch mass at $z-1$, and the mass of all the major branch progenitors (including the major branch itself) 
at $z$, i.e. $M_{sf}(z,z-1) = M_{*,MB}(z-1) - M_{*,allprogMB}(z)$, where $M_{*,allprogMB}$ 
is the stellar mass in all the progenitors of the major branch. 

We note here that while the `mass growth due to star formation' $M_{sf}$ so calculated is an accurate estimate
of the amount of new stellar mass contributed to the final $z \simeq 6$ galaxy in the preceding $\Delta z = 1$,
we cannot be sure that all this star-formation takes place `within' the major branch. 
Some of this star formation could clearly
take place in the minor progenitors before they finally merge with the major branch
at some point between $z$ and $z -1$. However, to better define what fraction of this 
star-formation activity takes place within the major branch would require the analysis of simulation snapshots very 
closely spaced in $z$; since such snapshots have not been stored, this calculation would 
require re-running the simulation which is not feasible at present. In any case, regardless 
of the precise location of the star-formation activity, this calculation 
delivers a robust estimate of the contribution of star formation in the preceding 
$\Delta z = 1$ to the stellar mass of the galaxy.

\begin{figure*} 
   \center{\includegraphics[scale=1.0]{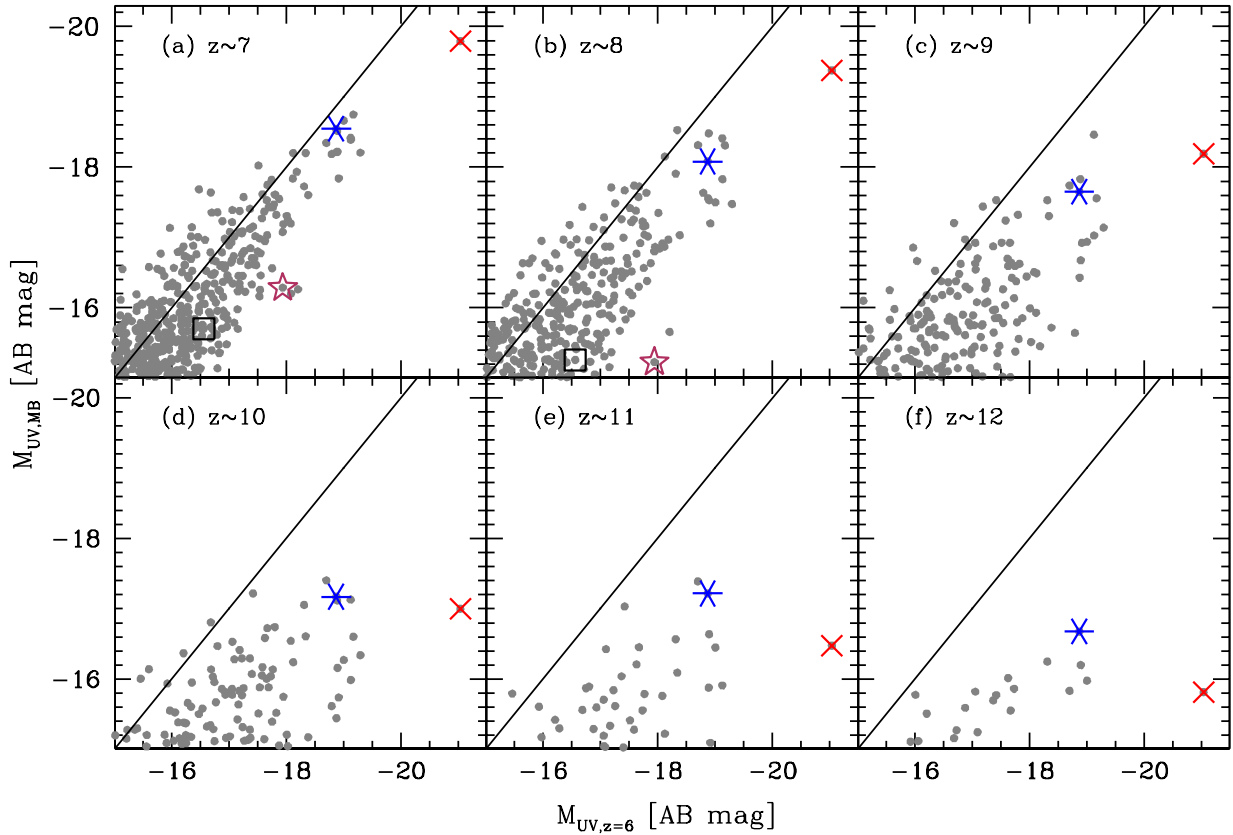}} 
  \caption{The absolute UV magnitude of the major branch progenitors with $M_{UV,MB}<-15$ for each $z\simeq 6$ LBG (small points), 
as a function of the final absolute UV magnitude at $z \simeq 6$. 
The panels show the results for different redshifts, $z\simeq 7-12$, as marked, 
and the large symbols show the results for the example $z \simeq 6$ LBGs discussed in the 
text: B (empty square), C (empty star), D (asterix) and E (cross); 
galaxy A does not appear on this plot since its progenitors are always fainter than the limiting value of $M_{UV}<-15$ shown here.}
\label{muvmb} 
\end{figure*}

The main results of these calculations are as follows. In all redshift intervals,
for the majority of galaxies,  
we find most of the mass growth results from star formation in the major branch, 
with mergers bringing in (relatively) tiny amounts of stellar mass such that 
$M_{sf}>>M_{acc}$ as can be seen from Fig. \ref{rat_sfmer} and Table 1; 
indeed, as is seen from the same table, $\simeq 90$\% of the total stellar mass of $z \simeq 6$ 
LBGs is assembled by star formation in the major branch, with mergers contributing 
only $\simeq 10$\% to this final value. Nevertheless, while this average behaviour is clear, 
there is a wide variety in the history of individual galaxies, and at almost all $z$, 
several galaxies in fact show values of $M_{acc} \approx M_{sf}$. Our result that the bulk of $M_{*,z=6}$ has to be built in the 
$\simeq 150$\,Myr between $z \simeq 7$ and $z \simeq 6$ is verified by panel (a) of Fig. \ref{rat_sfmer} 
where star formation in the major branch leads to an average of about 56\% of $M_{*,z=6}$ being formed between these 
redshifts. Finally, we note that since high-$z$ LBGs are expected to evolve into massive early-type galaxies at low redshifts, our results are in excellent agreement with the `two-phase' scenario of galaxy formation, where galaxies grow rapidly in mass by `in-situ' star formation at high-redshifts ($z \gsim 3$) with minor mergers contributing considerably to the mass buildup at later times ($z \lsim 2-3$), as shown by \citet{naab2009}, \citet{ oser2010} and \citet{ johansson2012}.

\begin{figure*} 
   \center{\includegraphics[scale=1.0]{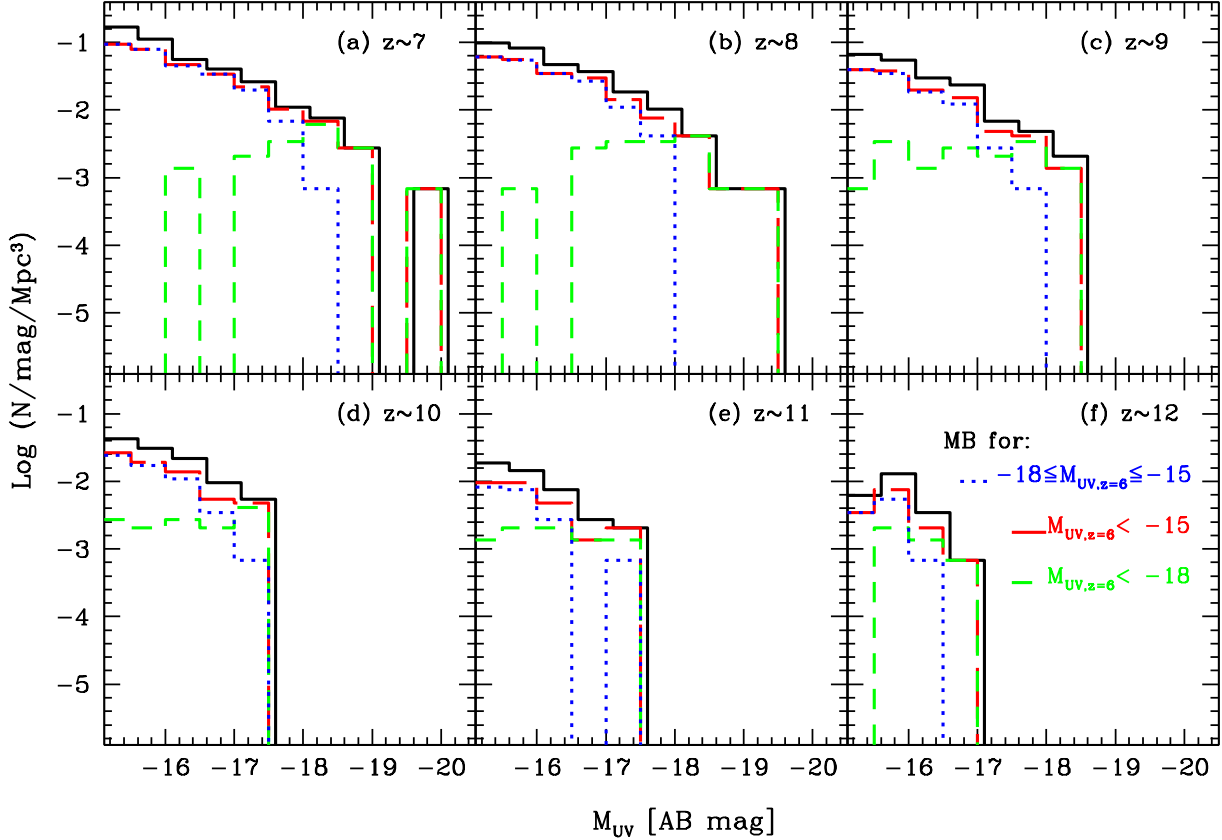}} 
  \caption{UV LFs for the major branch of the merger tree for $z\simeq6$ LBGs, for $z \simeq 7$ to $z \simeq 12$, as marked in each panel. 
In each panel, the long dashed (red), short dashed (green) and dotted (blue) lines show the major branch of the 
merger tree for $z \simeq 6$ LBGs with $M_{UV,z=6}< -15$, $M_{UV,z=6}< -18$ and $-18\leq M_{UV,z=6} \leq -15$, 
respectively. For comparison, the solid (black) line in each panel shows the UV LF obtained 
using all the galaxies in the $10{\rm cMpc}$ simulation box with $M_{UV}<-15$, 
at the redshift marked in that panel. In all panels, we have slightly displaced the solid black line horizontally for clearer visualization.}
\label{mb_uvlf} 
\end{figure*}

We can try to further clarify these results by again discussing the histories of our five example galaxies: 
as shown in Fig. \ref{msmb}, galaxies D, E have built up only about $3,0.3$\% of $M_{*,z=6}$ by $z \simeq 11$; 
from panel (f) of Fig. \ref{rat_sfmer}, we can clearly see that while for E this mass has solely been built up by $M_{sf}$, 
for D, there is some significant contribution due to $M_{acc}$. These galaxies then slowly build up 
their stellar mass, mostly through star formation until $z \simeq 9-8, 8-7$ when mergers of smaller systems 
start making a noticeable contribution; mergers contribute as much to the stellar 
mass as star formation in the major branch for C, D and E at $z \simeq 8-6$, $8-7$ and $9-8$, respectively. 
On the other hand, galaxy E only has a single progenitor for most of its life and thus 
grows solely by star formation in the major branch, except between $z \simeq 8$ and $z \simeq 7$ when 
it has two progenitors (see also fig. \ref{rat_mball}) and merger of the other major-branch 
progenitor contributes a small amount to $M_{*,z=6}$. Finally, about 40-70\% of $M_{*,z=6}$ 
for these five example galaxies is built-up by star formation in the $\simeq 150$\,Myr between $z \simeq 7$ and $z\simeq 6$, with mergers contributing only about 2-25\% 
to $M_{*,z=6}$ (the remaining mass pre-existing in the major branch at $z \simeq 7$).  

\section{UV luminosity evolution}
\label{lum_assembly}
Now that we have discussed how $z \simeq 6$ LBGs assemble their stellar mass, we can also study 
how their UV luminosity varies over time. 
We address this question by comparing the absolute UV magnitude of the major branch 
progenitor ($M_{UV,MB}$) at each redshift, with the final absolute UV magnitude of the 
$z \simeq 6$ LBGs, $M_{UV,z=6}$, as shown in Fig. \ref{muvmb}. 
Although $\langle E(B-V) \rangle<0.05$ at $z \simeq 6$ 
and is expected to decrease further with increasing $z$ as a result of galaxies being 
younger and less metal enriched, we have nonetheless calculated the dust 
enrichment of each galaxy in all the snapshots used between $z \simeq 12$ and $z \simeq 6$, 
consistently applying the dust model described in Sec. \ref{identifying_lbgs}. 

The first result we find is that $M_{UV,MB} \propto M_{UV,z=6}$ at least for 
$z \simeq 7-10$, consistent with the most massive $z \simeq 6$ LBGs having 
the most massive major branch progenitor; 
this correlation becomes sketchy at $z \geq 11$ as a result of 
the scarcity of progenitors brighter than the applied magnitude limit of $M_{UV}<-15$ 
above this redshift. The average absolute magnitude shifts towards lower values 
(i.e. increasing luminosity) with decreasing $z$, thus producing the effect of luminosity 
evolution in the luminosity function (see below). Moreover, the brightest sources appear to display fairly steady 
luminosity evolution on a source-by-source basis; for example, the progenitors of the 
$z \simeq 6$ galaxies with $M_{UV} < -18.5$ constitute $\simeq 90$\% of the galaxies 
with $M_{UV} < -18$ at $z \simeq 7$, and still provide $\simeq 75$\% of the 
galaxies brighter than $M_{UV} \simeq -18$ at $z \simeq 8$. 

However, at fainter luminosities the situation is more complex, as illustrated by the fact that between $z \simeq 7$
and $z \simeq 10$ a number of individual major-branch progenitors with $M_{*,z=6}<10^{8}\Msun$ are more luminous 
than their (higher mass) $z \simeq 6$ descendants (i.e. $M_{UV,MB} <M_{UV,z=6}$, as shown in panels (a) to (d) 
of Fig. \ref{muvmb}). This result may seem puzzling, given that even at $z \simeq 7$, these small galaxies have only 
been able to assemble a stellar mass $M_{*,MB}\lsim 0.4 M_{*,z=6}$, but is explained by the more stochastic star-formation 
histories of the lower-mass galaxies (as illustrated in Fig. \ref{sfr_fnz} in the appendix). In other words, 
a short-lived burst of star-formation in a low mass major-branch progenitor, which decays on times-scales of a few Myr, 
can easily produce a temporary enhancement in UV luminosity which exceeds that of the (still fairly low mass) descendant at 
$z \simeq 6$. The stochastic star formation in low mass galaxies is possibly the effect of these tiny potential wells losing a substantial part/all of their gas in SN driven outflows, suppressing further star formation; these progenitors must then wait for enough gas to be accreted, either from the IGM or through mergers, to re-ignite star formation. Obviously, the star formation rates in these galaxies are dependent on the mechanical feedback model used and changing the model parameters such as the mass upload rate and the fraction of SN energy carried away by winds could alter our results: increasing the mass upload rate (fraction of SN energy carried away by winds) would lead to a decrease (increase) in the wind velocity as it leaves the galactic disk, leading to these galaxies retaining (losing) more of their star forming gas. 

As for the 5 example galaxies that have been discussed above, the smallest at $z \simeq 6$, galaxy A is 
right at the limit of our magnitude cut, with $M_{UV,z=6} =-15.1$, and its major-branch progenitors are never luminous enough
to clear this threshold at higher redshift. The major branches of B and C ($M_{UV,z=6} \simeq -16.6,-17.9$) 
already have a stellar mass $M_{*,MB} \simeq (15, 6)\% M_{*,z=6}$ respectively, 
when they become visible with $M_{UV,MB} \simeq -15.2$ at $z \simeq 8$. 
For D and E, which have final $z \simeq 6$ magnitudes of $M_{UV,z=6} \simeq -18.8,-21$ 
respectively, the major-branch progenitors are visible at all the redshifts $z \simeq 7-12$ with the UV magnitude decreasing (i.e. 
their luminosities brightening) monotonically
with decreasing $z$. However, we note that the UV luminosity of E grows faster compared to D 
for $z \leq 9$ as a result of its faster stellar mass growth that naturally produces more UV photons, consistent 
with our finding above that most mass growth in a given redshift interval is driven by star formation. 

Finally, we show, and deconstruct the simulated UV LFs for LBGs in the simulation with $M_{UV}<-15$, over the redshift 
range $z \simeq 7-12$ (see also Table 2). The full simulated UV LFs at $z \simeq 6,7,8$ have already been shown in Fig. \ref{fig_lbg_uvlf},
but in Fig. \ref{mb_uvlf} we extend this to predict the form of the total UV LF up to $z \simeq 12$, and also separate out the 
contribution of the major-branch progenitors to the LF. At all the redshifts shown ($z \simeq 7-12$), 
the major branch progenitors of the $z \simeq 6$ LBGs dominate the UV LF over the entire magnitude range probed here. At 
the brightest magnitudes they are, not surprisingly, completely dominant, and even at the faint end ($M_{UV}\gsim -16$)
they constitute $\simeq 60$\% of the number density of objects in the LF. 

\begin{table*} 
\begin{center} 
\caption {For the redshifts shown in column 1, we show the total number of simulated LBGs in our simulated box of size $(10 {\rm cMpc})^3$ that contain at least 10 star particles with $M_{UV}<-15$ (column 2), with $-16<M_{UV}<-15$ (column 3), $-17<M_{UV}<-16$ (column 4), $-18<M_{UV}<-17$ (column 5), $-19<M_{UV}<-18$ (column 6) and $-19<M_{UV}<-20$ (column 7). As seen from Fig. \ref{res_uvlf}, while increasing the selection criterion to a minimum of 20 star particles does not affect the UV LFs at any redshift, increasing this value to 50 star particles only leads to a drop in the number of galaxies in the narrow range between $M_{UV}=-15 $ and $-15.5$ for $z \simeq 7-12$; with the drop increasing with increasing redshift.   }
\begin{tabular}{|c|c|c|c|c|c|c} 
\hline 
$z$ & $LBG_T$ & $LBG_{15-16}$ & $LBG_{16-17}$ & $LBG_{17-18}$ & $LBG_{18-19}$ & $LBG_{19-20}$\\
$$ & $$ & $$ & $$ & $$ & $$ & $$\\ 
\hline
$6$ & $871$ & $585$ & $197$ & $66$ & $18$ & $5$ \\
$7$ & $621$ & $410 $ & $141$ & $54$ & $15$ & $1$ \\
$8$ & $434$ & $262$ & $122$ & $42$ & $7$ & $1$ \\
$9$ & $273$ & $176$ & $77 $ & $17$ & $3$ & $0$ \\
$10$ & $161$ & $107 $ & $46$ & $8$ & $0$ & $0$ \\
$11$ & $66$ & $48$ & $15 $ & $3$ & $0 $ & $0$ \\
$12$ & $34 $ & $28$ & $6 $ & $0$ & $0$ & $0$ \\
\hline
\label{table2} 
\end{tabular} 
\end{center}
\end{table*} 

To further clarify the physical evolution of the galaxy population, we have also produced the UV LF for the major-branch 
progenitors of $z \simeq 6$ LBGs with (a) $M_{UV,z=6}<-18$ and (b) $-18 \leq M_{UV,z=6}\leq -15$. We find that the major-branch 
progenitors of the former dominate the high luminosity end of the UV LF at all $z$ and make a negligible 
contribution to the faint end. By contrast, the progenitors of the faintest $z \simeq 6$ LBGs with $-18 \leq M_{UV,z=6}\leq -15$ 
contribute to the faint end of the UV LF at all the redshifts studied. 

In conclusion, our simulation suggests that it is indeed reasonable to expect a steady brightening of the bright end of 
UV LF during the first billion years, and that this is primarily driven by {\it genuine physical luminosity evolution}
(i.e. steady brightening, albeit not exponential) of a fixed subset of the highest-mass LBGs. However, at fainter magnitudes the situation is 
clearly much more complex, involving a mix of positive and negative luminosity evolution (as low-mass galaxies temporarily
brighten then fade) coupled with both positive and negative density evolution (as new low-mass galaxies 
form, and other low-mass galaxies are consumed by merging).

\section{Conclusions}
\label{conc}
We have used state-of-the-art high-resolution SPH cosmological simulations, 
specially crafted to include the physics most relevant to galaxy formation 
(star formation, gas cooling and feedback) and a new treatment for metal enrichment 
and its dispersion that allows us to study the transition from metal free PopIII to metal enriched PopII star formation, 
and hence simulate the emergence of the first generations of galaxies in the high-redshift universe.
By combining simulation snapshots at $z \simeq 6-12$ with a previously developed dust model \citep{dayal2010a}, 
in addition to calculating the LBG UV LFs, $SMD$ and $sSFR$ for the current magnitude limit of $M_{UV} \simeq -18$, 
we have extended our results down to $M_{UV} \simeq -15$ in order to make specific predictions for upcoming instruments 
such as the JWST. The main results from our comparison of the population predictions from the theoretical model and 
the constraints provided by current data can be summarized as follows.

\begin{enumerate}

\item{{\it UV LFs:} we find that the simulated LBG UV LFs match both the amplitude and 
the slope of the observations in the current range of overlap; 
the faint-end slope is found to remain almost constant between $z \simeq 6$ and $z \simeq 8$, with a value consistent with $\alpha \simeq -2$.}

\item{{\it SMD:} the $SMD$ for the simulated LBGs with $M_{UV}<-18$ is in extremely good agreement with the observed values at $z \simeq 6-8$. 
Interestingly, we find that as a result of their much larger number density, about 1.5 times the currently-observed stellar 
mass is locked up in the undetected, faint LBGs with $ -18 \leq M_{UV} \leq -15$ at these redshifts.}

\item{{\it sSFR:} we find that the $sSFR$ decreases gently with increasing $M_*$ values since even a small amount of star formation in 
galaxies with low $M_*$ values is enough to push up the $sSFR$. The theoretical $sSFR$ increases from about $4.5\, {\rm Gyr}^{-1}$ at $z \simeq 6$ 
to about $11\, {\rm Gyr}^{-1}$ at $z \simeq 9$. In the currently accessible redshift range of overlap ($z \simeq 6-7$), 
the simulated sSFRs are in excellent agreement with those inferred from the latest observations (especially after 
correcting for emission-line contributions), but the above figures obviously provide a strong prediction 
of rapidly increasing typical values of $sSFR$ at higher redshifts, 
a prediction  potentially testable with the JWST.}

\end{enumerate}

Having demonstrated that the population predictions from the simulations 
match well with the existing observational data, we have validated our model as a potentially viable 
description of the physical growth and evolution of high-redshift galaxies. We have therefore proceeded
to examine the detailed galaxy-by-galaxy behaviour within the simulation in an attempt to gain 
physical insight into the observed population statistics. 
To do this we have identified all the progenitors as well as the major branch of the merger tree 
for $z \simeq 6$ LBGs ($M_{UV}<-15$) at higher redshift $z \simeq 7,8,9,10,11,12$. This has
enabled us to study how these $z \simeq 6$ LBGs have assembled their stellar mass over the preceding 
$\simeq 600$\,Myr of cosmic history, and also investigate how their UV luminosity has varied during this time.

The story that emerges regarding the lives of these galaxies can now be summarized as follows.
At $z \simeq 12$, the progenitors of $z \simeq 6$ LBGs contain, on average, only 0.3\% of the 
final $z \simeq 6$ stellar mass and the rate of stellar-mass build-up increases with decreasing 
redshift such that by $z \simeq 7$, these progenitors contain about 37\% of the total stellar mass 
contained in $z \simeq 6$ LBGs; this increasing SF efficiency appears to be the result of 
the decreasing effects of negative feedback (blowout/blowaway) as the potential well of the 
major branch grows. These numbers imply that the bulk ($\simeq 56$\%) of $M_*$ is formed in the 
$\simeq 150$\,Myr between $z \simeq 7$ and $z \simeq 6$. Further, we find, unsurprisingly, that the progenitors 
of the most massive $z \simeq 6$ LBGs start assembling first; 
while progenitors of $z \simeq 6$ LBGs with $M_* >10^8 \Msun$ exist as early as 
$z \simeq 12$, progenitors of the lowest mass ($M_* \leq 10^7 \Msun$) $z \simeq 6$ galaxies appear as late as 
$z \simeq 9$. As a result of their early assembly, the most massive $z \simeq 6$ LBGs generally 
have the most massive major branch progenitor at the redshifts studied. We briefly digress here to caution the reader that the results regarding the assembly of high-$z$ galaxies, specially in their early stages will depend on the feedback model used: in the simulation used in this work, galactic winds have a fixed velocity of $500 \, {\rm km \, s^{-1}}$ with a mass upload rate equal to twice the local star formation rate, and carry away a fixed fraction ($25\%$) of the SN energy ($10^{51} \,{\rm ergs}$). Increasing the mass upload rate (fraction of SN energy carried away by winds) would lead to a decrease (increase) in the wind velocity as it leaves the galactic disk, leading to these galaxies retaining (losing) more of their star forming gas. Qualitatively, this would result in smaller (larger) amounts of stellar mass having been assembled in both the major branch and all the other progenitors. However, quantifying the relative importance of varying the feedback parameters on the major branch and the other progenitors at different redshifts would require re-running the simulation by varying the feedback parameters, which is much beyond the scope of this work.

Compared to all the other progenitors (i.e. excluding the major branch itself), 
the major branch is always the dominant $M_*$ component; at $z \simeq 12$ 
the major branch typically contains about twice the total mass across all other progenitors of $z \simeq 6$ 
LBGs, and by $z \simeq 7$ this ratio has risen to about six. 
In the redshift range $z \simeq 12-10$, the major branch and all the other progenitors grow at approximately 
the same rate, but thereafter the major branch grows 
more quickly in $M_*$, possibly due to the decreasing impact of negative feedback.
However, while the global trend is clear, at every redshift there are outliers in which the major branch contains 
as little as 10\% of the total stellar mass assembled by that redshift, highlighting the varied histories 
of the galaxies in the simulation. 

We have also attempted to determine the relative importance of mass added by pre-existing stars in merging progenitors, 
as compared to that added by new star-formation activity in a given redshift interval.
To compute this we have defined the stellar mass `accreted' by the major branch at redshift $z-1$ as the stellar mass 
of all its other progenitors at $z$ and the mass gained by `star formation' as the difference of the stellar mass of the major branch 
at $z-1$ and all its progenitors at $z$. From this calculation, we have shown that $\simeq 90$\% of the 
total $z \simeq 6$ stellar mass is built up by star formation, and mergers contribute only $\simeq 10$\% to the final stellar mass. 
However, the average contribution of mergers increases with decreasing redshift, 
from $\simeq 0.01$\% at $z \simeq 12$ to  $\simeq 6$\% at $z \simeq 6$, 
as more and more progenitors form and fall into the ever-increasing major-branch potential well. 

Finally, we have tracked the variation of UV luminosity in the galaxies and their various progenitors
over the redshift range $z \simeq 12 - 6$. At least since $z \simeq 9$, we find that it is largely the same set of 
relatively massive galaxies which occupy the bright end of the UV luminosity function, and that these objects
brighten steadily with time (albeit not exponentially), thus providing a physical basis for interpreting the evolution
of the bright end of the LF in terms of luminosity evolution.
However, at fainter luminosities the situation is more complex, as illustrated by the fact that between $z \simeq 7$
and $z \simeq 10$ a number of individual major-branch progenitors with $M_{*,z=6}<10^{8}\Msun$ are more luminous
than their (higher mass) $z \simeq 6$ descendants; this appears to be one consequence of the more stochastic 
star-formation histories of the lower-mass galaxies in our simulation (as illustrated by the examples of individual 
galaxy formation histories provided in the appendix).

In conclusion, our simulation suggests that it is indeed reasonable to expect a steady brightening of the bright end of
UV LF during the first billion years, and that this is primarily driven by {\it genuine physical luminosity evolution}
(i.e. steady brightening) of a fixed subset of the highest-mass LBGs. However, at fainter magnitudes the situation is
clearly much more complex, involving a mix of positive and negative luminosity evolution (as low-mass galaxies temporarily
brighten then fade) coupled with both positive and negative density evolution (as new low-mass galaxies
form, and other low-mass galaxies are consumed by merging). Thus, it is probably not reasonable to aim to
explain the entire evolution of the UV LF, over a broad range in absolute magnitude, in terms of either pure
luminosity evolution or pure density evolution. However, improved data are clearly required to distinguish between
the viability of the model presented here, and alternative scenarios.

\section*{Acknowledgments} 
JSD and PD acknowledge the support of the European Research Council via the award of an Advanced Grant, and JSD also acknowledges the support of the Royal Society via a Wolfson Research Merit award. PD thanks A. Ferrara for his insightful comments, E.R. Tittley for his invaluable help in post-processing the simulation snapshots, and E. Curtis-Lake and R.J. McLure for their invaluable help in linking the theoretical results to the observations. UM acknowledges financial contribution from the HPC-Europa2 grant 228398 with the support of the European Community, under the FP7 Research Infrastructure Programme, and funding from a Marie Curie fellowship of the European Union Seventh Framework Programme (FP7/2007-2013) under grant agreement n. 267251.


\bibliographystyle{mn2e}
\bibliography{lfmv}

\begin{thebibliography}{110}
\expandafter\ifx\csname natexlab\endcsname\relax\def\natexlab#1{#1}\fi

\bibitem[{{Abel} {et~al}\mbox{.}(2002){Abel}, {Bryan}, \& {Norman}}]{abel2002}
{Abel} T., {Bryan} G.~L., {Norman} M.~L., 2002, Science, 295, 93

\bibitem[{{Barkana} \& {Loeb}(2001)}]{barkana-loeb2001}
{Barkana} R., {Loeb} A., 2001, \physrep, 349, 125

\bibitem[{{Bate} \& {Burkert}(1997)}]{bate-burkert1997}
{Bate} M.~R., {Burkert} A., 1997, \mnras, 288, 1060

\bibitem[{{Bianchi} \& {Schneider}(2007)}]{bianchi-schneider2007}
{Bianchi} S., {Schneider} R., 2007, \mnras, 378, 973

\bibitem[{{Blumenthal} {et~al}\mbox{.}(1984){Blumenthal}, {Faber}, {Primack},
  \& {Rees}}]{blumenthal1984}
{Blumenthal} G.~R., {Faber} S.~M., {Primack} J.~R., {Rees} M.~J., 1984, \nat,
  311, 517

\bibitem[{{Bouwens} {et~al}\mbox{.}(2007){Bouwens}, {Illingworth}, {Franx}, \&
  {Ford}}]{bouwens2007}
{Bouwens} R.~J., {Illingworth} G.~D., {Franx} M., {Ford} H., 2007, \apj, 670,
  928

\bibitem[{{Bouwens} {et~al}\mbox{.}(2010{\natexlab{a}}){Bouwens},
  {Illingworth}, {Gonz{\'a}lez}, {Labb{\'e}}, {Franx}, {Conselice},
  {Blakeslee}, {van Dokkum}, {Holden}, {Magee}, {Marchesini}, \&
  {Zheng}}]{bouwens2010a}
{Bouwens} R.~J. {et~al.}, 2010{\natexlab{a}}, \apj, 725, 1587

\bibitem[{{Bouwens} {et~al}\mbox{.}(2012){Bouwens}, {Illingworth}, {Oesch},
  {Franx}, {Labb{\'e}}, {Trenti}, {van Dokkum}, {Carollo}, {Gonz{\'a}lez},
  {Smit}, \& {Magee}}]{bouwens2012}
{Bouwens} R.~J. {et~al.}, 2012, \apj, 754, 83

\bibitem[{{Bouwens} {et~al}\mbox{.}(2011){Bouwens}, {Illingworth}, {Oesch},
  {Labb{\'e}}, {Trenti}, {van Dokkum}, {Franx}, {Stiavelli}, {Carollo},
  {Magee}, \& {Gonzalez}}]{bouwens2011}
{Bouwens} R.~J. {et~al.}, 2011, \apj, 737, 90

\bibitem[{{Bouwens} {et~al}\mbox{.}(2010{\natexlab{b}}){Bouwens},
  {Illingworth}, {Oesch}, {Trenti}, {Stiavelli}, {Carollo}, {Franx}, {van
  Dokkum}, {Labb{\'e}}, \& {Magee}}]{bouwens2010}
{Bouwens} R.~J. {et~al.}, 2010{\natexlab{b}}, \apjl, 708, L69

\bibitem[{{Bouwens} {et~al}\mbox{.}(2004){Bouwens}, {Thompson}, {Illingworth},
  {Franx}, {van Dokkum}, {Fan}, {Dickinson}, {Eisenstein}, \&
  {Rieke}}]{bouwens2004}
{Bouwens} R.~J. {et~al.}, 2004, \apjl, 616, L79

\bibitem[{{Bowler} {et~al}\mbox{.}(2012){Bowler}, {Dunlop}, {McLure},
  {McCracken}, {Milvang-Jensen}, {Furusawa}, {Fynbo}, {Le Fevre}, {Holt},
  {Ideue}, {Ihara}, {Rogers}, \& {Taniguchi}}]{bowler2012}
{Bowler} R.~A.~A. {et~al.}, 2012, ArXiv e-prints

\bibitem[{{Bradley} {et~al}\mbox{.}(2012){Bradley}, {Trenti}, {Oesch},
  {Stiavelli}, {Treu}, {Bouwens}, {Shull}, {Holwerda}, \&
  {Pirzkal}}]{bradley2012}
{Bradley} L.~D. {et~al.}, 2012, ArXiv e-prints

\bibitem[{{Bromm} \& {Loeb}(2003)}]{bromm-loeb2003}
{Bromm} V., {Loeb} A., 2003, \nat, 425, 812

\bibitem[{{Bromm} {et~al}\mbox{.}(2009){Bromm}, {Yoshida}, {Hernquist}, \&
  {McKee}}]{bromm2009}
{Bromm} V., {Yoshida} N., {Hernquist} L., {McKee} C.~F., 2009, \nat, 459, 49

\bibitem[{{Bunker} {et~al}\mbox{.}(2010){Bunker}, {Wilkins}, {Ellis}, {Stark},
  {Lorenzoni}, {Chiu}, {Lacy}, {Jarvis}, \& {Hickey}}]{bunker2010}
{Bunker} A.~J. {et~al.}, 2010, \mnras, 409, 855

\bibitem[{{Campisi} {et~al}\mbox{.}(2011){Campisi}, {Maio}, {Salvaterra}, \&
  {Ciardi}}]{campisi2011}
{Campisi} M.~A., {Maio} U., {Salvaterra} R., {Ciardi} B., 2011, \mnras, 416,
  2760

\bibitem[{{Castellano} {et~al}\mbox{.}(2010{\natexlab{a}}){Castellano},
  {Fontana}, {Boutsia}, {Grazian}, {Pentericci}, {Bouwens}, {Dickinson},
  {Giavalisco}, {Santini}, {Cristiani}, {Fiore}, {Gallozzi}, {Giallongo},
  {Maiolino}, {Mannucci}, {Menci}, {Moorwood}, {Nonino}, {Paris}, {Renzini},
  {Rosati}, {Salimbeni}, {Testa}, \& {Vanzella}}]{castellano2010b}
{Castellano} M. {et~al.}, 2010{\natexlab{a}}, \aap, 511, A20

\bibitem[{{Castellano} {et~al}\mbox{.}(2010{\natexlab{b}}){Castellano},
  {Fontana}, {Paris}, {Grazian}, {Pentericci}, {Boutsia}, {Santini}, {Testa},
  {Dickinson}, {Giavalisco}, {Bouwens}, {Cuby}, {Mannucci}, {Cl{\'e}ment},
  {Cristiani}, {Fiore}, {Gallozzi}, {Giallongo}, {Maiolino}, {Menci},
  {Moorwood}, {Nonino}, {Renzini}, {Rosati}, {Salimbeni}, \&
  {Vanzella}}]{castellano2010a}
{Castellano} M. {et~al.}, 2010{\natexlab{b}}, \aap, 524, A28

\bibitem[{{Choudhury} \& {Ferrara}(2007)}]{choudhury-ferrara2007}
{Choudhury} T.~R., {Ferrara} A., 2007, \mnras, 380, L6

\bibitem[{{Ciardi} \& {Ferrara}(2005)}]{ciardi-ferrara2005}
{Ciardi} B., {Ferrara} A., 2005, \ssr, 116, 625

\bibitem[{{Curtis-Lake} {et~al}\mbox{.}(2012{\natexlab{a}}){Curtis-Lake},
  {McLure}, {Dunlop}, {Schenker}, {Rogers}, {Targett}, {Cirasuolo}, {Almaini},
  {Ashby}, {Bradshaw}, {Finkelstein}, {Dickinson}, {Ellis}, {Faber}, {Fazio},
  {Ferguson}, {Fontana}, {Grogin}, {Hartley}, {Kocevski}, {Koekemoer}, {Lai},
  {Robertson}, {Vanzella}, \& {Willner}}]{curtis-lake2012b}
{Curtis-Lake} E. {et~al.}, 2012{\natexlab{a}}, ArXiv e-prints

\bibitem[{{Curtis-Lake} {et~al}\mbox{.}(2012{\natexlab{b}}){Curtis-Lake},
  {McLure}, {Pearce}, {Dunlop}, {Cirasuolo}, {Stark}, {Almaini}, {Bradshaw},
  {Chuter}, {Foucaud}, \& {Hartley}}]{curtis-lake2012a}
{Curtis-Lake} E. {et~al.}, 2012{\natexlab{b}}, \mnras, 422, 1425

\bibitem[{{Daddi} {et~al}\mbox{.}(2007){Daddi}, {Dickinson}, {Morrison},
  {Chary}, {Cimatti}, {Elbaz}, {Frayer}, {Renzini}, {Pope}, {Alexander},
  {Bauer}, {Giavalisco}, {Huynh}, {Kurk}, \& {Mignoli}}]{daddi2007}
{Daddi} E. {et~al.}, 2007, \apj, 670, 156

\bibitem[{{Dayal} \& {Ferrara}(2012)}]{dayal2012}
{Dayal} P., {Ferrara} A., 2012, \mnras, 421, 2568

\bibitem[{{Dayal} {et~al}\mbox{.}(2010){Dayal}, {Ferrara}, \&
  {Saro}}]{dayal2010a}
{Dayal} P., {Ferrara} A., {Saro} A., 2010, \mnras, 402, 1449

\bibitem[{{Dayal} {et~al}\mbox{.}(2009){Dayal}, {Ferrara}, {Saro},
  {Salvaterra}, {Borgani}, \& {Tornatore}}]{dayal2009}
{Dayal} P., {Ferrara} A., {Saro} A., {Salvaterra} R., {Borgani} S., {Tornatore}
  L., 2009, \mnras, 400, 2000

\bibitem[{{Dayal} {et~al}\mbox{.}(2011){Dayal}, {Maselli}, \&
  {Ferrara}}]{dayal2011a}
{Dayal} P., {Maselli} A., {Ferrara} A., 2011, \mnras, 410, 830

\bibitem[{{Dolag} {et~al}\mbox{.}(2009){Dolag}, {Borgani}, {Murante}, \&
  {Springel}}]{dolag2009}
{Dolag} K., {Borgani} S., {Murante} G., {Springel} V., 2009, \mnras, 399, 497

\bibitem[{{Dunlop}(2012)}]{dunlop2012b}
{Dunlop} J.~S., 2012, ArXiv e-prints

\bibitem[{{Dunlop} {et~al}\mbox{.}(2012){Dunlop}, {McLure}, {Robertson},
  {Ellis}, {Stark}, {Cirasuolo}, \& {de Ravel}}]{dunlop2012}
{Dunlop} J.~S., {McLure} R.~J., {Robertson} B.~E., {Ellis} R.~S., {Stark}
  D.~P., {Cirasuolo} M., {de Ravel} L., 2012, \mnras, 420, 901

\bibitem[{{Dwek} {et~al}\mbox{.}(2007){Dwek}, {Galliano}, \&
  {Jones}}]{dwek2007}
{Dwek} E., {Galliano} F., {Jones} A.~P., 2007, Nuovo Cimento B Serie, 122, 959

\bibitem[{{Ferrara} \& {Loeb}(2012)}]{ferrara2012}
{Ferrara} A., {Loeb} A., 2012, ArXiv e-prints

\bibitem[{{Ferrara} {et~al}\mbox{.}(2000){Ferrara}, {Pettini}, \&
  {Shchekinov}}]{ferrara2000}
{Ferrara} A., {Pettini} M., {Shchekinov} Y., 2000, \mnras, 319, 539

\bibitem[{{Finkelstein} {et~al}\mbox{.}(2010){Finkelstein}, {Papovich},
  {Giavalisco}, {Reddy}, {Ferguson}, {Koekemoer}, \&
  {Dickinson}}]{finkelstein2010}
{Finkelstein} S.~L., {Papovich} C., {Giavalisco} M., {Reddy} N.~A., {Ferguson}
  H.~C., {Koekemoer} A.~M., {Dickinson} M., 2010, \apj, 719, 1250

\bibitem[{{Finkelstein} {et~al}\mbox{.}(2012{\natexlab{a}}){Finkelstein},
  {Papovich}, {Ryan}, {Pawlik}, {Dickinson}, {Ferguson}, {Finlator},
  {Koekemoer}, {Giavalisco}, {Cooray}, {Dunlop}, {Faber}, {Grogin}, {Kocevski},
  \& {Newman}}]{finkelstein2012b}
{Finkelstein} S.~L. {et~al.}, 2012{\natexlab{a}}, \apj, 758, 93

\bibitem[{{Finkelstein} {et~al}\mbox{.}(2012{\natexlab{b}}){Finkelstein},
  {Papovich}, {Salmon}, {Finlator}, {Dickinson}, {Ferguson}, {Giavalisco},
  {Koekemoer}, {Reddy}, {Bassett}, {Conselice}, {Dunlop}, {Faber}, {Grogin},
  {Hathi}, {Kocevski}, {Lai}, {Lee}, {McLure}, {Mobasher}, \&
  {Newman}}]{finkelstein2012a}
{Finkelstein} S.~L. {et~al.}, 2012{\natexlab{b}}, \apj, 756, 164

\bibitem[{{Finlator} {et~al}\mbox{.}(2007){Finlator}, {Dav{\'e}}, \&
  {Oppenheimer}}]{finlator2007}
{Finlator} K., {Dav{\'e}} R., {Oppenheimer} B.~D., 2007, \mnras, 376, 1861

\bibitem[{{Forero-Romero} {et~al}\mbox{.}(2011){Forero-Romero}, {Yepes},
  {Gottl{\"o}ber}, {Knollmann}, {Cuesta}, \& {Prada}}]{forero2011}
{Forero-Romero} J.~E., {Yepes} G., {Gottl{\"o}ber} S., {Knollmann} S.~R.,
  {Cuesta} A.~J., {Prada} F., 2011, \mnras, 415, 3666

\bibitem[{{Gonz{\'a}lez} {et~al}\mbox{.}(2011){Gonz{\'a}lez}, {Labb{\'e}},
  {Bouwens}, {Illingworth}, {Franx}, \& {Kriek}}]{gonzalez2011}
{Gonz{\'a}lez} V., {Labb{\'e}} I., {Bouwens} R.~J., {Illingworth} G., {Franx}
  M., {Kriek} M., 2011, \apjl, 735, L34

\bibitem[{{Gonz{\'a}lez} {et~al}\mbox{.}(2010){Gonz{\'a}lez}, {Labb{\'e}},
  {Bouwens}, {Illingworth}, {Franx}, {Kriek}, \& {Brammer}}]{gonzalez2010}
{Gonz{\'a}lez} V., {Labb{\'e}} I., {Bouwens} R.~J., {Illingworth} G., {Franx}
  M., {Kriek} M., {Brammer} G.~B., 2010, \apj, 713, 115

\bibitem[{{Greif} {et~al}\mbox{.}(2011){Greif}, {Springel}, {White}, {Glover},
  {Clark}, {Smith}, {Klessen}, \& {Bromm}}]{greif2011}
{Greif} T.~H., {Springel} V., {White} S.~D.~M., {Glover} S.~C.~O., {Clark}
  P.~C., {Smith} R.~J., {Klessen} R.~S., {Bromm} V., 2011, \apj, 737, 75

\bibitem[{{Haardt} \& {Madau}(1996)}]{haardt-madau1996}
{Haardt} F., {Madau} P., 1996, \apj, 461, 20

\bibitem[{{Heger} \& {Woosley}(2002)}]{heger-woosley2002}
{Heger} A., {Woosley} S.~E., 2002, \apj, 567, 532

\bibitem[{{Iliev} {et~al}\mbox{.}(2012){Iliev}, {Mellema}, {Shapiro}, {Pen},
  {Mao}, {Koda}, \& {Ahn}}]{iliev2012}
{Iliev} I.~T., {Mellema} G., {Shapiro} P.~R., {Pen} U.-L., {Mao} Y., {Koda} J.,
  {Ahn} K., 2012, \mnras, 423, 2222

\bibitem[{{Johansson} {et~al}\mbox{.}(2012){Johansson}, {Naab}, \&
  {Ostriker}}]{johansson2012}
{Johansson} P.~H., {Naab} T., {Ostriker} J.~P., 2012, \apj, 754, 115

\bibitem[{{Komatsu} {et~al}\mbox{.}(2009){Komatsu}, {Dunkley}, {Nolta},
  {Bennett}, {Gold}, {Hinshaw}, {Jarosik}, {Larson}, {Limon}, {Page},
  {Spergel}, {Halpern}, {Hill}, {Kogut}, {Meyer}, {Tucker}, {Weiland},
  {Wollack}, \& {Wright}}]{komatsu2009}
{Komatsu} E. {et~al.}, 2009, \apjs, 180, 330

\bibitem[{{Labb{\'e}} {et~al}\mbox{.}(2010{\natexlab{a}}){Labb{\'e}},
  {Gonz{\'a}lez}, {Bouwens}, {Illingworth}, {Franx}, {Trenti}, {Oesch}, {van
  Dokkum}, {Stiavelli}, {Carollo}, {Kriek}, \& {Magee}}]{labbe2010a}
{Labb{\'e}} I. {et~al.}, 2010{\natexlab{a}}, \apjl, 716, L103

\bibitem[{{Labb{\'e}} {et~al}\mbox{.}(2010{\natexlab{b}}){Labb{\'e}},
  {Gonz{\'a}lez}, {Bouwens}, {Illingworth}, {Oesch}, {van Dokkum}, {Carollo},
  {Franx}, {Stiavelli}, {Trenti}, {Magee}, \& {Kriek}}]{labbe2010b}
{Labb{\'e}} I. {et~al.}, 2010{\natexlab{b}}, \apjl, 708, L26

\bibitem[{{Labbe} {et~al}\mbox{.}(2012){Labbe}, {Oesch}, {Bouwens},
  {Illingworth}, {Magee}, {Gonzalez}, {Carollo}, {Franx}, {Trenti}, {van
  Dokkum}, \& {Stiavelli}}]{labbe2012}
{Labbe} I. {et~al.}, 2012, ArXiv e-prints

\bibitem[{{Larson}(1998)}]{larson1998}
{Larson} R.~B., 1998, \mnras, 301, 569

\bibitem[{{Leitherer} {et~al}\mbox{.}(1999){Leitherer}, {Schaerer}, {Goldader},
  {Gonz{\'a}lez Delgado}, {Robert}, {Kune}, {de Mello}, {Devost}, \&
  {Heckman}}]{leitherer1999}
{Leitherer} C. {et~al.}, 1999, APJS, 123, 3

\bibitem[{{Lorenzoni} {et~al}\mbox{.}(2011){Lorenzoni}, {Bunker}, {Wilkins},
  {Stanway}, {Jarvis}, \& {Caruana}}]{lorenzoni2011}
{Lorenzoni} S., {Bunker} A.~J., {Wilkins} S.~M., {Stanway} E.~R., {Jarvis}
  M.~J., {Caruana} J., 2011, \mnras, 414, 1455

\bibitem[{{Maeder} \& {Meynet}(1989)}]{maeder-meynet1989}
{Maeder} A., {Meynet} G., 1989, \aap, 210, 155

\bibitem[{{Maio} {et~al}\mbox{.}(2010){Maio}, {Ciardi}, {Dolag}, {Tornatore},
  \& {Khochfar}}]{maio2010}
{Maio} U., {Ciardi} B., {Dolag} K., {Tornatore} L., {Khochfar} S., 2010,
  \mnras, 407, 1003

\bibitem[{{Maio} {et~al}\mbox{.}(2009){Maio}, {Ciardi}, {Yoshida}, {Dolag}, \&
  {Tornatore}}]{maio2009}
{Maio} U., {Ciardi} B., {Yoshida} N., {Dolag} K., {Tornatore} L., 2009, \aap,
  503, 25

\bibitem[{{Maio} {et~al}\mbox{.}(2007){Maio}, {Dolag}, {Ciardi}, \&
  {Tornatore}}]{maio2007}
{Maio} U., {Dolag} K., {Ciardi} B., {Tornatore} L., 2007, \mnras, 379, 963

\bibitem[{{Maio} {et~al}\mbox{.}(2011){Maio}, {Khochfar}, {Johnson}, \&
  {Ciardi}}]{maio2011}
{Maio} U., {Khochfar} S., {Johnson} J.~L., {Ciardi} B., 2011, \mnras, 414, 1145

\bibitem[{{Mannucci} {et~al}\mbox{.}(2007){Mannucci}, {Buttery}, {Maiolino},
  {Marconi}, \& {Pozzetti}}]{mannucci2007}
{Mannucci} F., {Buttery} H., {Maiolino} R., {Marconi} A., {Pozzetti} L., 2007,
  \aap, 461, 423

\bibitem[{{Martin}(1999)}]{martin1999}
{Martin} C.~L., 1999, \apj, 513, 156

\bibitem[{{McCracken} {et~al}\mbox{.}(2012){McCracken}, {Milvang-Jensen},
  {Dunlop}, {Franx}, {Fynbo}, {Le F{\`e}vre}, {Holt}, {Caputi}, {Goranova},
  {Buitrago}, {Emerson}, {Freudling}, {Hudelot}, {L{\'o}pez-Sanjuan},
  {Magnard}, {Mellier}, {M{\o}ller}, {Nilsson}, {Sutherland}, {Tasca}, \&
  {Zabl}}]{mccracken2012}
{McCracken} H.~J. {et~al.}, 2012, \aap, 544, A156

\bibitem[{{McKee}(1989)}]{mckee1989}
{McKee} C., 1989, in IAU Symposium, Vol. 135, Interstellar Dust, {Allamandola}
  L.~J., {Tielens} A.~G.~G.~M., eds., p. 431

\bibitem[{{McLure} {et~al}\mbox{.}(2009){McLure}, {Cirasuolo}, {Dunlop},
  {Foucaud}, \& {Almaini}}]{mclure2009}
{McLure} R.~J., {Cirasuolo} M., {Dunlop} J.~S., {Foucaud} S., {Almaini} O.,
  2009, \mnras, 395, 2196

\bibitem[{{McLure} {et~al}\mbox{.}(2010){McLure}, {Dunlop}, {Cirasuolo},
  {Koekemoer}, {Sabbi}, {Stark}, {Targett}, \& {Ellis}}]{mclure2010}
{McLure} R.~J., {Dunlop} J.~S., {Cirasuolo} M., {Koekemoer} A.~M., {Sabbi} E.,
  {Stark} D.~P., {Targett} T.~A., {Ellis} R.~S., 2010, \mnras, 403, 960

\bibitem[{{McLure} {et~al}\mbox{.}(2011){McLure}, {Dunlop}, {de Ravel},
  {Cirasuolo}, {Ellis}, {Schenker}, {Robertson}, {Koekemoer}, {Stark}, \&
  {Bowler}}]{mclure2011}
{McLure} R.~J. {et~al.}, 2011, \mnras, 418, 2074

\bibitem[{{McQuinn} {et~al}\mbox{.}(2007){McQuinn}, {Hernquist}, {Zaldarriaga},
  \& {Dutta}}]{mcquinn2007}
{McQuinn} M., {Hernquist} L., {Zaldarriaga} M., {Dutta} S., 2007, \mnras, 381,
  75

\bibitem[{{Naab} {et~al}\mbox{.}(2009){Naab}, {Johansson}, \&
  {Ostriker}}]{naab2009}
{Naab} T., {Johansson} P.~H., {Ostriker} J.~P., 2009, \apjl, 699, L178

\bibitem[{{Nagamine} {et~al}\mbox{.}(2010){Nagamine}, {Ouchi}, {Springel}, \&
  {Hernquist}}]{nagamine2010}
{Nagamine} K., {Ouchi} M., {Springel} V., {Hernquist} L., 2010, \pasj, 62, 1455

\bibitem[{{Nakamura} \& {Umemura}(2001)}]{nakamura-umemura2001}
{Nakamura} F., {Umemura} M., 2001, \apj, 548, 19

\bibitem[{{Navarro} \& {White}(1993)}]{navarro-white1993}
{Navarro} J.~F., {White} S.~D.~M., 1993, \mnras, 265, 271

\bibitem[{{Neistein} \& {Dekel}(2008)}]{neistein2008}
{Neistein} E., {Dekel} A., 2008, \mnras, 383, 615

\bibitem[{{Nozawa} {et~al}\mbox{.}(2007){Nozawa}, {Kozasa}, {Habe}, {Dwek},
  {Umeda}, {Tominaga}, {Maeda}, \& {Nomoto}}]{nozawa2007}
{Nozawa} T., {Kozasa} T., {Habe} A., {Dwek} E., {Umeda} H., {Tominaga} N.,
  {Maeda} K., {Nomoto} K., 2007, \apj, 666, 955

\bibitem[{{Oesch} {et~al}\mbox{.}(2010){Oesch}, {Bouwens}, {Illingworth},
  {Carollo}, {Franx}, {Labb{\'e}}, {Magee}, {Stiavelli}, {Trenti}, \& {van
  Dokkum}}]{oesch2010}
{Oesch} P.~A. {et~al.}, 2010, \apjl, 709, L16

\bibitem[{{Oesch} {et~al}\mbox{.}(2012){Oesch}, {Bouwens}, {Illingworth},
  {Gonzalez}, {Trenti}, {van Dokkum}, {Franx}, {Labbe}, {Carollo}, \&
  {Magee}}]{oesch2012}
{Oesch} P.~A. {et~al.}, 2012, ArXiv e-prints

\bibitem[{{Omukai} \& {Palla}(2003)}]{omukai-palla2003}
{Omukai} K., {Palla} F., 2003, \apj, 589, 677

\bibitem[{{Oser} {et~al}\mbox{.}(2010){Oser}, {Ostriker}, {Naab}, {Johansson},
  \& {Burkert}}]{oser2010}
{Oser} L., {Ostriker} J.~P., {Naab} T., {Johansson} P.~H., {Burkert} A., 2010,
  \apj, 725, 2312

\bibitem[{{Ouchi} {et~al}\mbox{.}(2009){Ouchi}, {Mobasher}, {Shimasaku},
  {Ferguson}, {Fall}, {Ono}, {Kashikawa}, {Morokuma}, {Nakajima}, {Okamura},
  {Dickinson}, {Giavalisco}, \& {Ohta}}]{ouchi2009}
{Ouchi} M. {et~al.}, 2009, \apj, 706, 1136

\bibitem[{{Ouchi} {et~al}\mbox{.}(2010){Ouchi}, {Shimasaku}, {Furusawa},
  {Saito}, {Yoshida}, {Akiyama}, {Ono}, {Yamada}, {Ota}, {Kashikawa}, {Iye},
  {Kodama}, {Okamura}, {Simpson}, \& {Yoshida}}]{ouchi2010}
{Ouchi} M. {et~al.}, 2010, \apj, 723, 869

\bibitem[{{Padovani} \& {Matteucci}(1993)}]{padovani-matteucci1993}
{Padovani} P., {Matteucci} F., 1993, \apj, 416, 26

\bibitem[{{Robertson} {et~al}\mbox{.}(2010){Robertson}, {Ellis}, {Dunlop},
  {McLure}, \& {Stark}}]{robertson2010}
{Robertson} B.~E., {Ellis} R.~S., {Dunlop} J.~S., {McLure} R.~J., {Stark}
  D.~P., 2010, \nat, 468, 49

\bibitem[{{Salpeter}(1955)}]{salpeter1955}
{Salpeter} E.~E., 1955, \apj, 121, 161

\bibitem[{{Salvaterra} {et~al}\mbox{.}(2011){Salvaterra}, {Ferrara}, \&
  {Dayal}}]{salvaterra2011}
{Salvaterra} R., {Ferrara} A., {Dayal} P., 2011, \mnras, 414, 847

\bibitem[{{Schaerer}(2002)}]{schaerer2002}
{Schaerer} D., 2002, \aap, 382, 28

\bibitem[{{Schaerer} \& {de Barros}(2009)}]{schaerer2009}
{Schaerer} D., {de Barros} S., 2009, \aap, 502, 423

\bibitem[{{Schenker} {et~al}\mbox{.}(2012){Schenker}, {Stark}, {Ellis},
  {Robertson}, {Dunlop}, {McLure}, {Kneib}, \& {Richard}}]{schenker2012}
{Schenker} M.~A., {Stark} D.~P., {Ellis} R.~S., {Robertson} B.~E., {Dunlop}
  J.~S., {McLure} R.~J., {Kneib} J.-P., {Richard} J., 2012, \apj, 744, 179

\bibitem[{{Schneider} {et~al}\mbox{.}(2003){Schneider}, {Ferrara},
  {Salvaterra}, {Omukai}, \& {Bromm}}]{schneider2003}
{Schneider} R., {Ferrara} A., {Salvaterra} R., {Omukai} K., {Bromm} V., 2003,
  \nat, 422, 869

\bibitem[{{Schneider} {et~al}\mbox{.}(2006){Schneider}, {Omukai}, {Inoue}, \&
  {Ferrara}}]{schneider2006}
{Schneider} R., {Omukai} K., {Inoue} A.~K., {Ferrara} A., 2006, \mnras, 369,
  1437

\bibitem[{{Seab} \& {Shull}(1983)}]{seab-shull1983}
{Seab} C.~G., {Shull} J.~M., 1983, \apj, 275, 652

\bibitem[{{Sheth} \& {Tormen}(1999)}]{sheth-tormen1999}
{Sheth} R.~K., {Tormen} G., 1999, \mnras, 308, 119

\bibitem[{{Springel}(2005)}]{springel2005}
{Springel} V., 2005, \mnras, 364, 1105

\bibitem[{{Springel} \& {Hernquist}(2003)}]{springel-hernquist2003b}
{Springel} V., {Hernquist} L., 2003, \mnras, 339, 289

\bibitem[{{Springel} {et~al}\mbox{.}(2001){Springel}, {Yoshida}, \&
  {White}}]{springel2001}
{Springel} V., {Yoshida} N., {White} S.~D.~M., 2001, \na, 6, 79

\bibitem[{{Stark} {et~al}\mbox{.}(2009){Stark}, {Ellis}, {Bunker}, {Bundy},
  {Targett}, {Benson}, \& {Lacy}}]{stark2009}
{Stark} D.~P., {Ellis} R.~S., {Bunker} A., {Bundy} K., {Targett} T., {Benson}
  A., {Lacy} M., 2009, \apj, 697, 1493

\bibitem[{{Stark} {et~al}\mbox{.}(2012){Stark}, {Schenker}, {Ellis},
  {Robertson}, {McLure}, \& {Dunlop}}]{stark2012}
{Stark} D.~P., {Schenker} M.~A., {Ellis} R.~S., {Robertson} B., {McLure} R.,
  {Dunlop} J., 2012, ArXiv e-prints

\bibitem[{{Suda} \& {Fujimoto}(2010)}]{suda-fujimoto2010}
{Suda} T., {Fujimoto} M.~Y., 2010, \mnras, 405, 177

\bibitem[{{Sutherland} \& {Dopita}(1993)}]{sutherland-dopita1993}
{Sutherland} R.~S., {Dopita} M.~A., 1993, \apjs, 88, 253

\bibitem[{{Thielemann} {et~al}\mbox{.}(2003){Thielemann}, {Argast},
  {Brachwitz}, {Hix}, {H{\"o}flich}, {Liebend{\"o}rfer}, {Martinez-Pinedo},
  {Mezzacappa}, {Panov}, \& {Rauscher}}]{thielemann2003}
{Thielemann} F.-K. {et~al.}, 2003, Nuclear Physics A, 718, 139

\bibitem[{{Todini} \& {Ferrara}(2001)}]{todini-ferrara2001}
{Todini} P., {Ferrara} A., 2001, \mnras, 325, 726

\bibitem[{{Tornatore} {et~al}\mbox{.}(2007{\natexlab{a}}){Tornatore},
  {Borgani}, {Dolag}, \& {Matteucci}}]{tornatore2007a}
{Tornatore} L., {Borgani} S., {Dolag} K., {Matteucci} F., 2007{\natexlab{a}},
  \mnras, 382, 1050

\bibitem[{{Tornatore} {et~al}\mbox{.}(2007{\natexlab{b}}){Tornatore},
  {Ferrara}, \& {Schneider}}]{tornatore2007b}
{Tornatore} L., {Ferrara} A., {Schneider} R., 2007{\natexlab{b}}, \mnras, 382,
  945

\bibitem[{{Valiante} {et~al}\mbox{.}(2009){Valiante}, {Matteucci}, {Recchi}, \&
  {Calura}}]{valiante2009}
{Valiante} R., {Matteucci} F., {Recchi} S., {Calura} F., 2009, \na, 14, 638

\bibitem[{{van den Hoek} \& {Groenewegen}(1997)}]{vandenhoek1997}
{van den Hoek} L.~B., {Groenewegen} M.~A.~T., 1997, \aaps, 123, 305

\bibitem[{{Weinmann} {et~al}\mbox{.}(2011){Weinmann}, {Neistein}, \&
  {Dekel}}]{weinmann2011}
{Weinmann} S.~M., {Neistein} E., {Dekel} A., 2011, \mnras, 417, 2737

\bibitem[{{Wilkins} {et~al}\mbox{.}(2011){Wilkins}, {Bunker}, {Stanway},
  {Lorenzoni}, \& {Caruana}}]{wilkins2011}
{Wilkins} S.~M., {Bunker} A.~J., {Stanway} E., {Lorenzoni} S., {Caruana} J.,
  2011, \mnras, 417, 717

\bibitem[{{Woosley} \& {Weaver}(1995)}]{woosley-weaver1995}
{Woosley} S.~E., {Weaver} T.~A., 1995, \apjs, 101, 181

\bibitem[{{Yajima} {et~al}\mbox{.}(2012){Yajima}, {Li}, {Zhu}, \&
  {Abel}}]{yajima2012}
{Yajima} H., {Li} Y., {Zhu} Q., {Abel} T., 2012, \mnras, 424, 884

\bibitem[{{Yoshida} {et~al}\mbox{.}(2003){Yoshida}, {Abel}, {Hernquist}, \&
  {Sugiyama}}]{yoshida2003}
{Yoshida} N., {Abel} T., {Hernquist} L., {Sugiyama} N., 2003, \apj, 592, 645

\bibitem[{{Yoshida} {et~al}\mbox{.}(2004){Yoshida}, {Bromm}, \&
  {Hernquist}}]{yoshida2004}
{Yoshida} N., {Bromm} V., {Hernquist} L., 2004, \apj, 605, 579

\bibitem[{{Yoshida} {et~al}\mbox{.}(2007){Yoshida}, {Omukai}, \&
  {Hernquist}}]{yoshida2007}
{Yoshida} N., {Omukai} K., {Hernquist} L., 2007, \apjl, 667, L117

\bibitem[{{Zheng} {et~al}\mbox{.}(2010){Zheng}, {Cen}, {Trac}, \&
  {Miralda-Escud{\'e}}}]{zheng2010}
{Zheng} Z., {Cen} R., {Trac} H., {Miralda-Escud{\'e}} J., 2010, \apj, 716, 574

\end{thebibliography}

\appendix

\section{Effect of minimum number of star particles chosen}

\begin{figure*} 
   \center{\includegraphics[scale=1.0]{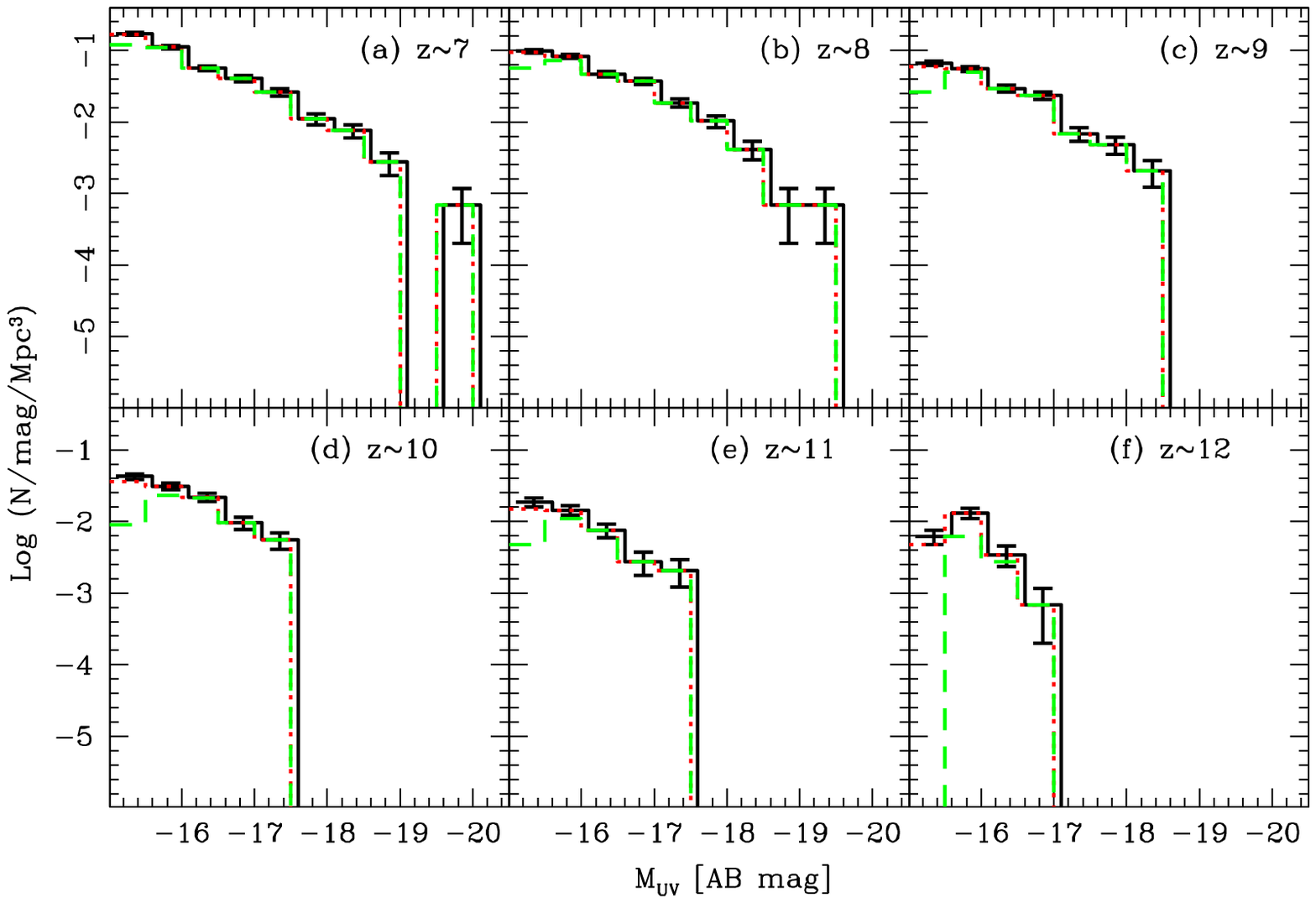}} 
  \caption{The UV LFs of galaxies at $z \simeq 7$ to $12$ as marked in each panel. 
In all panels, the solid, dotted and dashed histograms show the theoretical UV LFs for galaxies with $M_{UV}<-15$ containing a minimum of 10, 20 and 50 star particles, respectively. In all panels, we have slightly displaced the solid black line horizontally for clearer visualization.}
\label{res_uvlf} 
\end{figure*}

We remind the reader that each resolved galaxy used in our calculations fulfils three criterion: $M_h \geq 10^{8.5} \, \Msun$, more than $4N$ gas particles and a minimum of 10 star particles; increasing the minimum number of star particles used is likely to affect the selection of the least massive objects. To quantify this effect, we have carried out a brief resolution study, the results of which are shown in Fig. \ref{res_uvlf}. While increasing the selection criterion to a minimum of 20 star particles does not affect the UV LFs at any redshift, increasing this value to 50 star particles only leads to a drop in the number of galaxies in the narrow range between $M_{UV}=-15 $ and $-15.5$ for $z \simeq 7-12$; with the drop increasing with increasing redshift. 

\section{Examples of stellar mass assembly, SFR and sSFR for ${\bf z \simeq 6}$ LBGs}
\label{appendix}
We show the $z$-dependent stellar mass growth, and the $SFR$ and $sSFR$ evolution for the major branch of the merger tree for 
18 `typical' LBGs selected from the simulation at $z \simeq 6$, with $M_{*,z=6}$ ranging between $10^{7.0-9.8}\Msun$. 
In these illustrative plots, the $SFR$ has been calculated as $\dot M_* = [M_*(t)-M_*(t-1)]/\Delta t$ 
where $M_*(t) - M_*(t-1)$ shows the stellar mass built within a certain time step and $\Delta t$ 
is the width of the time-step. Note that this stellar mass includes both the mass built by star 
formation inside the major branch, as well as the mass gained by merging with progenitors of the 
major branch (see sec. \ref{sf_acc}). 

\begin{figure*} 
   \center{\includegraphics[scale=1.0]{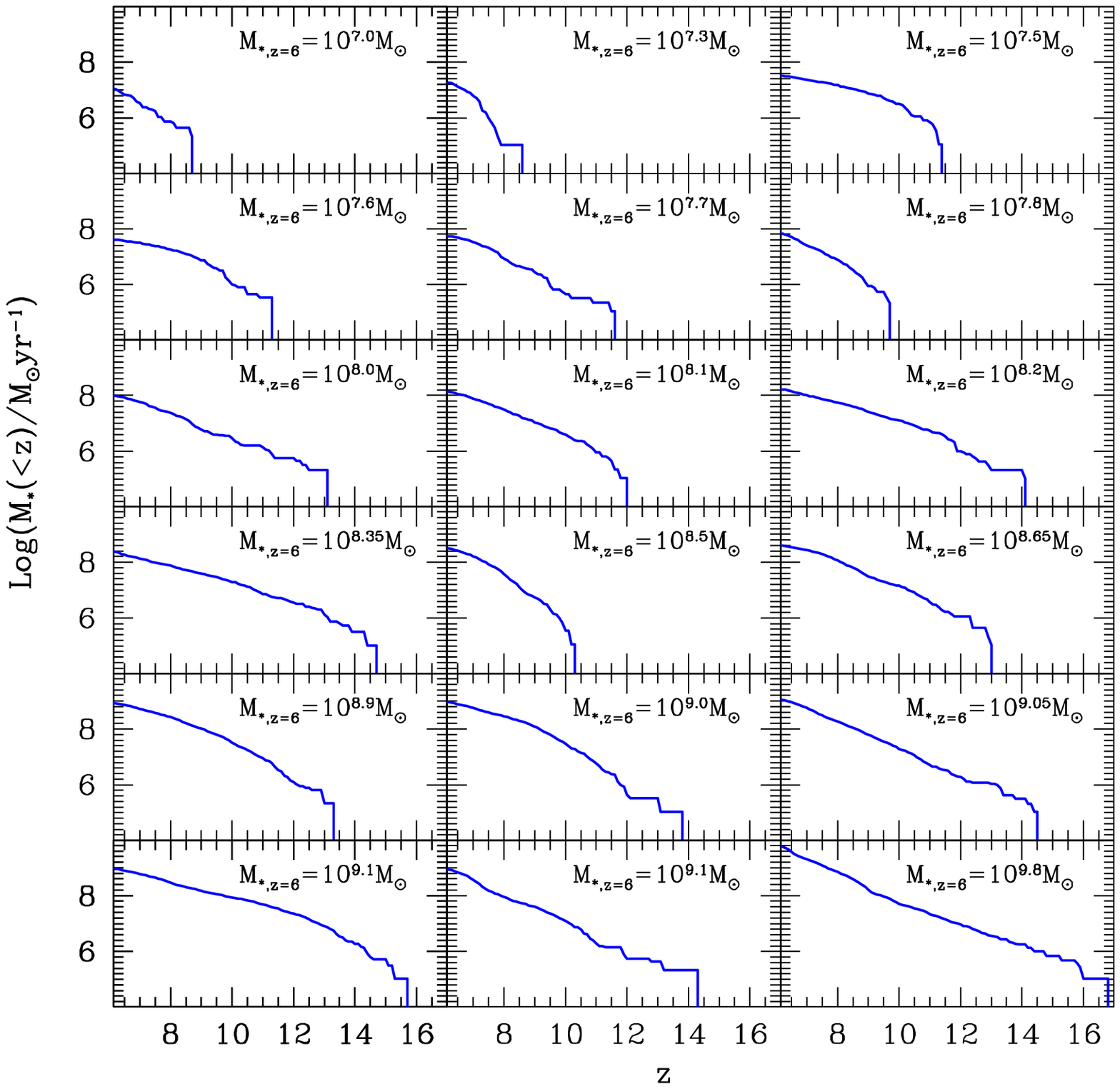}} 
  \caption{The total stellar mass assembled in the major branch by the redshift shown 
on the x-axis, either through mergers or star formation. Each of the panels shows the cumulative mass for different galaxies, 
with the $M_{*,z=6}$ value marked.}
\label{ms_fnz} 
\end{figure*}

\begin{figure*} 
   \center{\includegraphics[scale=1.0]{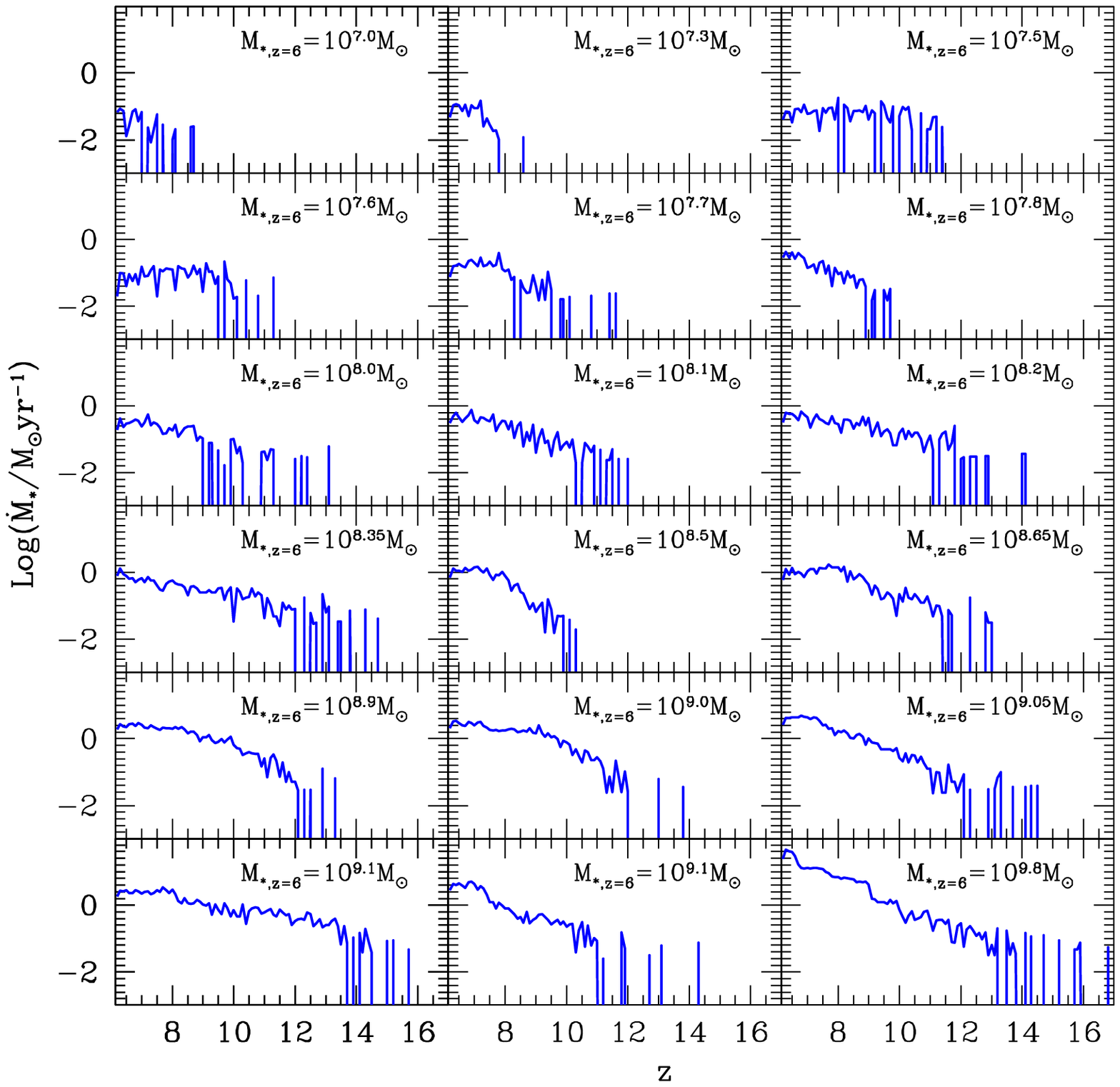}} 
  \caption{The $SFR$ computed as the ratio of the stellar mass assembled by the major branch 
in a given time step, by the time-step width, as a function of the redshift. 
Each of the panels shows the $SFR$ for different galaxies with the final $M_{*,z=6}$ value marked.}
\label{sfr_fnz} 
\end{figure*}

\begin{figure*}
   \center{\includegraphics[scale=1.0]{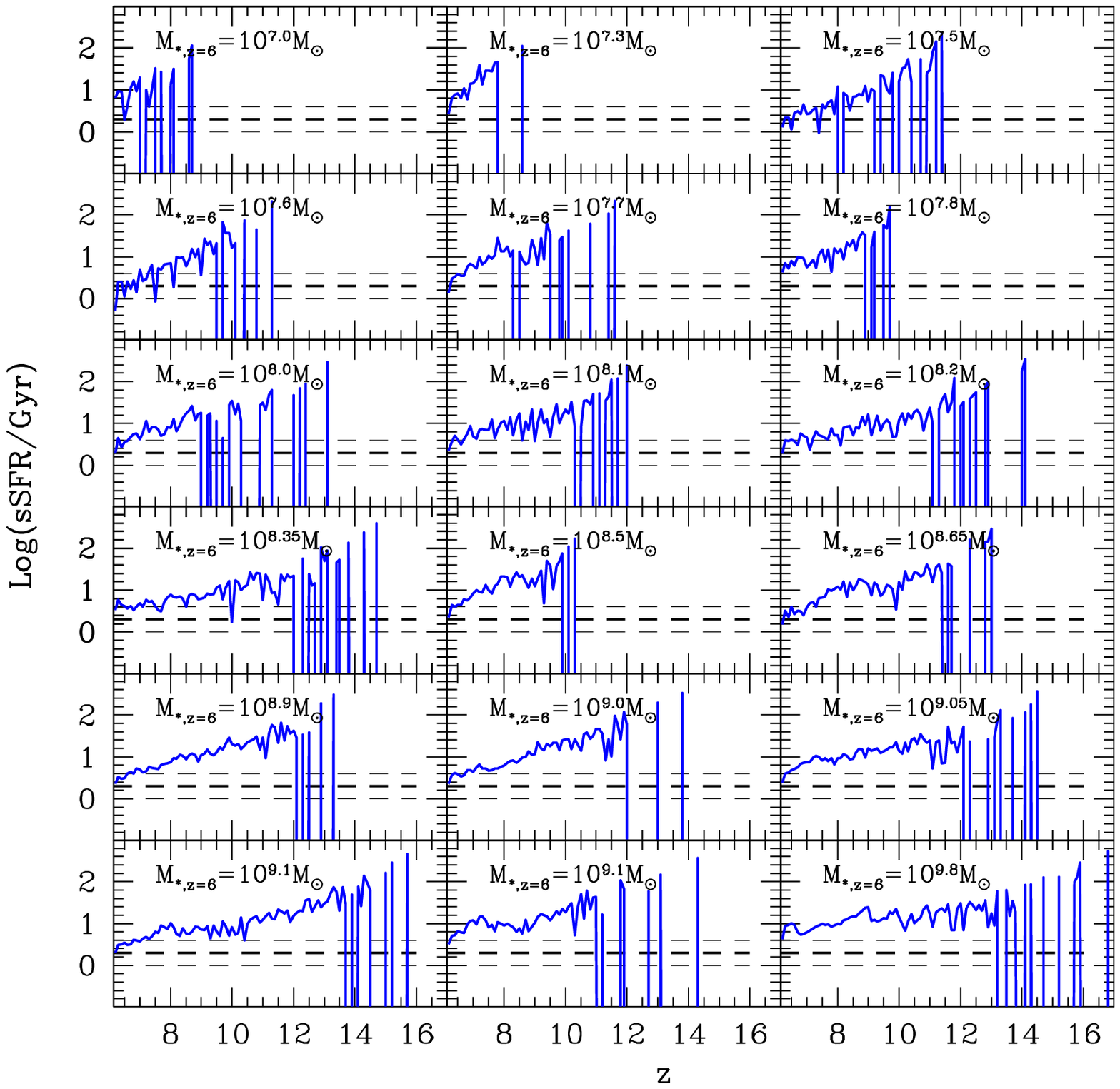}}
  \caption{The $sSFR$ computed as the ratio of the $SFR$ to the stellar mass assembled by the major branch, as a function of the redshift. 
Each of the panels shows the $sSFR$ for different galaxies with the final $M_{*,z=6}$ value marked. 
The horizontal thick (thin) lines show the average (spread) of the $sSFR$ computed observationally \citep{gonzalez2010} between $z \simeq 4$
and $z \simeq 8$; to guide the eye, these observational results have been extended to $z = 16$ assuming that the observed $sSFR$ remains constant to such redshifts.}
\label{ssfr_fnz}
\end{figure*}

\newpage 
\label{lastpage} 
\end{document}